\documentclass[10pt]{iopart}
\usepackage{iopams}
\usepackage{graphicx}
\usepackage{subfig}
\usepackage{amsmath, mathtools, amssymb}
\usepackage{textcomp}
\usepackage{placeins}
\usepackage{cases} 
\usepackage{xcolor}
\usepackage{hyperref}
\usepackage{cite}
\usepackage{caption}
\usepackage{float}

\usepackage{tabularx}
\usepackage{dblfloatfix}
\usepackage{comment}

\begin{document}
\title[]{Dynamics of reactive oxygen species produced by the COST microplasma jet} 

\author{Sascha Chur$^{1}$, Robin Minke$^{1}$, Youfan He$^{2}$, M\'{a}t\'{e} Vass$^{3,4}$, Thomas Mussenbrock$^{3}$, Ralf Peter Brinkmann$^{2}$, Efe Kemaneci$^{2}$, Lars Schücke$^{3}$, Volker Schulz-von der Gathen$^{5}$, Andrew R. Gibson$^{6,7}$, Marc Böke$^{5}$, Judith Golda$^{1}$}

\address{$^{1}$Plasma Interface Physics, Ruhr University Bochum, Bochum, Germany}
\address{$^{2}$Chair of Theoretical Electrical Engineering, Faculty of Electrical Engineering and Information Technology, Ruhr University Bochum, Bochum, Germany}
\address{$^{3}$Chair of Applied Electrodynamics and Plasma Technology, Faculty of Electrical Engineering and Information Technology, Ruhr University Bochum, Bochum, Germany}
\address{$^{4}$Institute for Solid State Physics and Optics, HUN-REN Wigner Research Centre for Physics, Budapest, Hungary}
\address{$^{5}$Experimental Physics II: Physics of Reactive Plasmas, Ruhr University Bochum, Bochum, Germany}
\address{$^{6}$Research Group for Biomedical Plasma Technology, Faculty of Electrical Engineering and Information Technology, Ruhr-University Bochum, Germany}
\address{$^{7}$York Plasma Institute, School of Physics, Engineering and Technology, University of York, United Kingdom}

\ead{sascha.chur@rub.de}
\vspace{10pt}
\begin{indented}
\item[January 2025]
\end{indented}

\begin{abstract}
This study is focused on measuring the densities of the excited molecular oxygen species, O$_{2}(\text{a}^{1}\Delta_{\text{g}})$ and O$_{2}(\text{b}^{1}\Sigma_{\text{g}}^{+})$, produced in a COST atmospheric pressure plasma jet using a helium-oxygen mixture. Knowledge of the ozone density is critical for measurements because of its high quenching rate of these species. Additionally O$_{2}(\text{a}^{1}\Delta_{\text{g}})$ is difficult to measure, due to its low emission intensity and sensitivity to background interference in the plasma region. Therefore a flow cell was used to enhance signal detection in the effluent region.
To validate the measurements and improve understanding of reaction mechanisms, results were compared with two simulation models: a pseudo-1D plug flow simulation and a 2D fluid simulation. The plug flow simulation provided an effective means for estimating species densities, with a fast computation time. The 2D simulation offered a more realistic description of the flow dynamics, which proved critical to correctly describe the experimental trends. However, it requires long computation times to reach an equilibrium state in the flow cell. Otherwise, it leads to discrepancies to the experimental data. Further discrepancies arose, from an overestimation of the ozone density from the models, as validated from the O$_{2}(\text{b}^{1}\Sigma_{\text{g}}^{+})$ density measurements. Optimizing the reaction rate coefficients for the effluent region might improve the agreement with the experimental results.
Despite these limitations both simulations aligned reasonably well with experimental data, showcasing the well validated plasma chemistry of the models, even for complicated effluent geometries.

\end{abstract}
\ioptwocol
%
%
%
%
%

\section{Introduction}

Micro atmospheric pressure plasma jets (µAPPJs)\cite{lu_atmospheric-pressure_2012,winter_atmospheric_2015,robert_characterization_2012,sobota_electric_2016,golda_concepts_2016,reuter_kinpenreview_2018} have garnered significant interest due to their wide-ranging applications, including surface treatment \cite{pawlat_rf_2016,shaw_mechanisms_2016}, etching \cite{ichiki_localized_2004}, and deposition \cite{benedikt_thin_2007,reuter_kinpenreview_2018}. These plasma sources offer several advantages over traditional methods. Notably, they have a simple setup since they do not require vacuum equipment, can be applied to heat and vacuum-sensitive surfaces, and their effluents can be easily used on various substrates. The efficacy of APPJs arises from their non-equilibrium discharge characteristics, where radio frequency (RF) excitation predominantly heats the electrons, while ions and neutrals remain relatively cool. \\

\noindent
Plasma jets often utilize argon or helium as carrier gas for stable plasma ignition, and are capable of producing reactive species through electron impact dissociation \cite{waskoenig_atomic_2010} when gases such as nitrogen \cite{preissing_three-dimensional_2020,bischoff_experimental_2018} and oxygen \cite{maletic_detection_2012,liu_micro_2021,korolov_atomic_2021} are added. However, the downstream reaction dynamics in these jets are complex, involving mechanisms like excitation, ionization, chemical reactions, dissociation, and quenching. Of particular interest are reactive oxygen and nitrogen species (RONS), which are crucial for biomedical applications such as cancer \cite{graves_reactive_2014,yang_role_2018} and wound treatment \cite{bekeschus_medical_2021,kong_plasma_2009}. Effective treatment relies on understanding the optimal dosage of reactive species, necessitating precise measurement of their densities and comprehension of their production mechanisms. \\

\noindent
Given the complexity of these interactions, modeling serves as a valuable tool for evaluating species interactions and densities, especially since experimental determination of several species' densities can be challenging. Zero-dimensional (0D) models provide a balance of accuracy and computational efficiency for simple gas mixtures like He/N$_{2}$ or He/O$_{2}$ \cite{lazzaroni_analyticalnumerical_2012,liu_global_2010,schroter_chemical_2018,schroter_numerical_2018,murakami_afterglow_2014,murakami_interacting_2013,murakami_chemical_2012,sun_global_2019,gaens_kinetic_2013,gaens_reaction_2014,gaens_numerical_2014,turner_uncertainty_2015,liu_main_2017,schmidt-bleker_plasma_2016,schmidt-bleker_reactive_2014}. 
These models can be used to study the influence of varying temperatures, gas mixture ratios, and plasma power on the plasma properties and reactive species production. For plasma jets, extending models to plug flow simulations \cite{schroter_chemical_2018,murakami_afterglow_2014,gaens_kinetic_2013,gaens_reaction_2014,gaens_numerical_2014} allows for spatial resolution of reactive species along the gas flow \cite{stafford_o_2004}, thereby describing time-dependent phenomena. \\

\noindent
If the effect of the discharge geometry becomes significant, for example, due to a non-trivial gas flow velocity distribution, or if details of the spatiotemporal plasma dynamics are required, more sophisticated simulation methods, such as fluid \cite{liu_1d_2020,kelly_gas_2015,viegas_physics_2022} or hybrid simulations \cite{vass_new_2024,klich_simulation_2022} are employed. \\

\noindent
Extensive studies have been conducted on various species, including reactive oxygen species \cite{murakami_chemical_2012,vass_new_2024,hemke_spatially_2011} as well as hydroxyl radicals \cite{schroter_chemical_2018}. Despite these efforts, a wide range of plasma properties still need to be benchmarked against experimental data to ensure the models' predictive capability. Due to limited diagnostic access to the micro-sized discharge regions of plasma jets, measuring many species remains challenging. \\

\noindent
The COST plasma jet \cite{golda_comparison_2020} offers a well-defined plasma source, making it an ideal candidate for simulation comparisons. Although it has been benchmarked against various simulations with good agreement to existing data \cite{vass_new_2024,klich_simulation_2022,he_zero-dimensional_2021}, many species' densities are yet to be validated. Further validation is essential to enhance understanding of the underlying reaction mechanisms, ultimately improving the effectiveness and reliability of atmospheric pressure plasma applications.\\

\noindent
This paper aims to advance the understanding of the reaction kinetics of oxygen species in the COST plasma jet by providing experimental data for difficult-to-detect species, like O$_{2}(\text{a}^{1}\Delta_{\text{g}})$ and O$_{2}(\text{b}^{1}\Sigma_{\text{g}}^{+})$, the first and second electronically excited states of molecular oxygen. Because they are heavily quenched by ozone, knowledge of its density is critical in calculating the densities of the excited molecular oxygen species. While O$_{2}(\text{b}^{1}\Sigma_{\text{g}}^{+})$ can be measured relatively easily by emission spectroscopy in the plasma region \cite{booth_quenching_2022}, but not in the effluent region, the determination of the O$_{2}(\text{a}^{1}\Delta_{\text{g}})$ density is more complicated. As a metastable state O$_{2}(\text{a}^{1}\Delta_{\text{g}})$ has a long lifetime with up to 75\,min \cite{ionin_physics_2007}, due to the forbidden optical transition of the singlet state. Its emission intensity at 1270\,nm is therefore low. Additionally, the thermal background in the plasma region makes it almost impossible to measure the emission signal, therefore mostly effluent measurements are feasible \cite{sousa_o_2008}. To enhance the signal in the effluent, gas flow cells are used to increase the observable volume \cite{sousa_o_2008}. Therefore this approach is used in our study. \\

\noindent
Here, the main focus lies on measuring the absolute densities of the excited molecular oxygen species O$_{2}(\text{a}^{1}\Delta_{\text{g}})$ and O$_{2}(\text{b}^{1}\Sigma_{\text{g}}^{+})$. The results were compared to two simulations, a pseudo-1D plug-flow simulation to benchmark the measured values as well as to provide insight into reaction dynamics by analysis of the simulation results, and a 2D fluid simulation to investigate the influence of the gas flow on the reaction dynamics. The distribution of the reactive species in the gas flow cell might have a significant influence on the measured densities. The pseudo-1D plug flow simulation, however, does not account for the spatial dynamics of the species inside the flow cell, which is inherently two-dimensional. For this purpose the 2D fluid simulation was employed, which not only provides spatio-temporally resolved species densities, but also offers insights into the flow velocity and gas temperature. However, this simulation method is computationally more expensive, which makes comparison of broad parameter variations time-consuming. For many applications, a rough estimate of the generated densities might be enough. For these cases the simple pseudo-1D plug flow simulation would suffice. This paper aims to use the combined insights of these methods to get a better understanding of the dynamics between ozone and the reactive species O$_{2}(\text{a}^{1}\Delta_{\text{g}})$ and O$_{2}(\text{b}^{1}\Sigma_{\text{g}}^{+})$. Furthermore, it tries to benchmark the oxygen species' densities and their interaction dynamics for the two commonly employed simulation models of the COST plasma jet, especially in the effluent region. Therefore, giving insight into the optimal use cases of the two simulations and their limitations.

\section{Methods}

\begin{figure}[H]
    \centering
    \includegraphics[width=\linewidth]{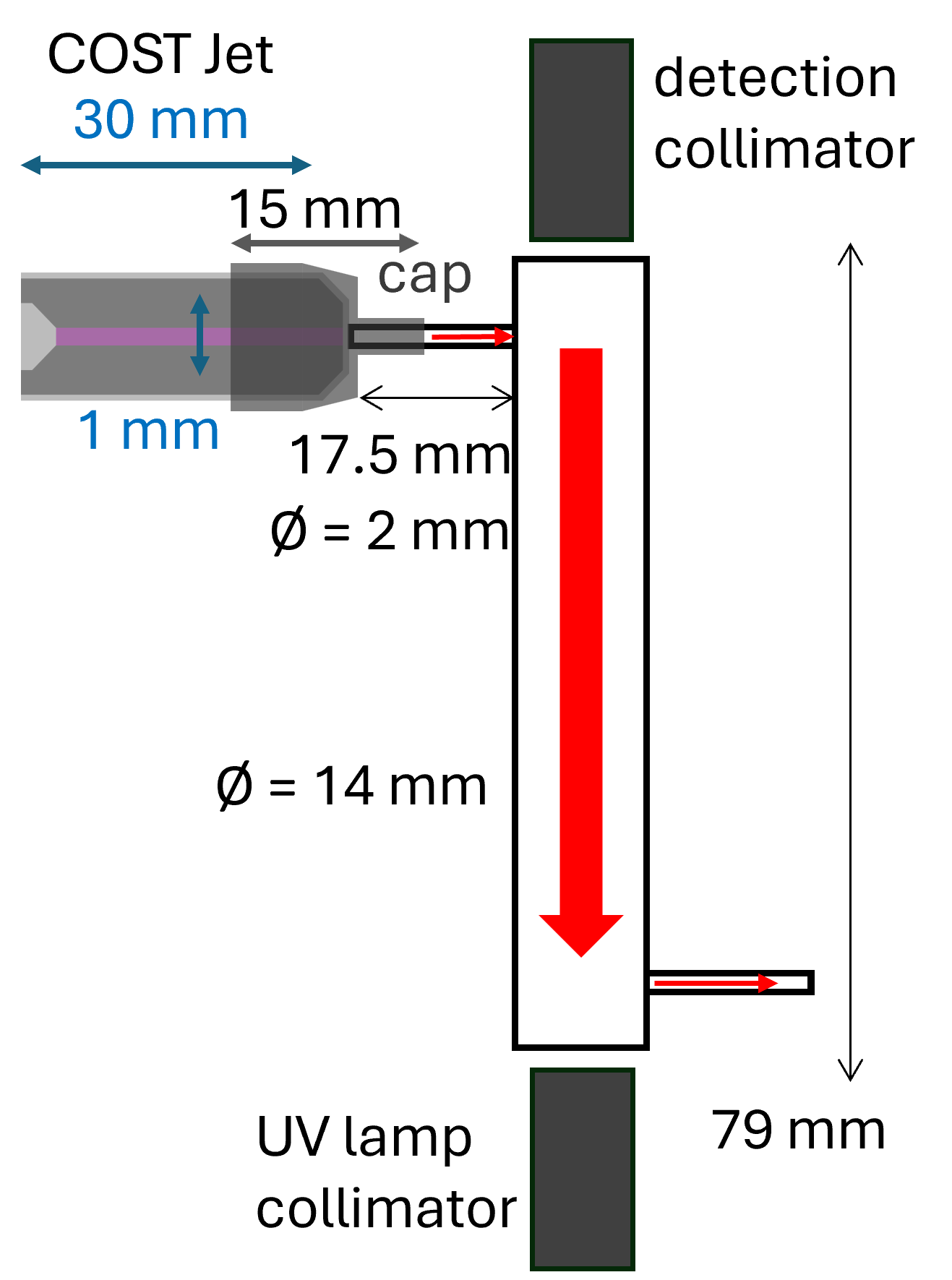}
    \caption{Schematic of the quartz flow cell dimensions and the connection to the COST jet, as well as the position of the collimators for the optical setups. The red arrows indicate the gas flow direction.}
    \label{fig:gas flow cell}
\end{figure}

\subsection{COST plasma jet}
The COST micro-scaled atmospheric pressure plasma jet, which is described in detail elsewhere \cite{golda_concepts_2016}, is well characterized and capable of generating a high density of reactive species. Its plasma chemistry is described by various validated simulations, allowing for reliable comparison with experimental values. It operates at a radio frequency of 13.56\,MHz using a plane electrode configuration. The discharge channel between the electrodes measures 1\,mm~×~1\,mm~×~30\,mm, with two quartz glass planes on the sides to provide optical access. Inside the jet, various reactive species can be produced and transported out of the active discharge volume, via the jet's nozzle. This gas stream is usually referred to as the effluent, and it can be used to treat surfaces with the produced chemical species. 

\subsection{Gas flow cell}\label{chp:flow cell}
The densities of many reactive species can be obtained by emission and absorption spectroscopy. These techniques can be challenging to apply in the effluent of a plasma jet because of the typically low optical emission, small observable volume, and short absorption length. To achieve a good signal-to-noise ratio for spectroscopic measurements, a large detection volume or long optical paths are favorable. This is especially critical for low emissivity from long-lived species like O$_{2}(\text{a}^{1}\Delta_{\text{g}})$. To this end, a gas flow cell was used to contain the effluent and increase the detection volume, in which light could be emitted or absorbed, respectively. The use of a gas flow cell does not allow for an analysis of the spatial distribution of species in the effluent region. However, it provides an estimate of the densities produced inside the plasma jet if the species is long-lived. The cells schematic is shown in Figure~\ref{fig:gas flow cell}. It is made out of quartz glass to allow the transmission of UV-light for absorption spectroscopy. The gas flow cell is connected to the COST plasma jet by a plastic cap, which guided the effluent into a 17.5\,mm long tube with a diameter of 2\,mm. The effluent then reaches the cylindrical gas flow cell with an inner diameter of 14\,mm and a length of 79\,mm. The plane windows at the ends of the flow cell allowed optical access. Attached at the outlet of the gas flow cell was a long pipe to prevent the backflow of air into the cell. \\

\subsection{Absorption spectroscopy of ozone}
The optical setup of the absorption spectroscopy measurement is shown in Figure \ref{fig:gas flow cell}. The utilized light source for the absorption spectroscopy was a laser driven broad band lamp (Energetiq EQ-99 LDLS). The light was guided through a fiber attached to the UV lamp and a collimator, which parallelized the light to a beam diameter of 13\,mm. This matched the gas flow cell dimensions. The light passed through a narrow bandpass UV filter (Edmund optics bandpass filter, OD 4, CW = 254\,nm, FWHM = 10\,nm). This wavelength range fits to the absorption of the Hartley band of ozone while reducing the energy input of the UV light which could lead to dissociation of O$_{2}$ and thus further ozone formation. The intensity of the absorption spectra was calculated from the averaged intensity within the filter wavelength range (approx. 250-260\,nm). At the end of the gas flow cell another collimator and optical fiber were used to guide the light into a UV sensitive spectrometer (Avantes AVASPEC-ULS-2048x64TEC-EVO, UB from 200\,nm to 720\,nm; resolution 0.3\,nm). \\

\noindent
To measure the density of O$_{2}(\text{a}^{1}\Delta_{\text{g}})$, knowledge of the ozone density is crucial due to its significant quenching effect on O$_{2}(\text{a}^{1}\Delta_{\text{g}})$. The ozone density was determined using absorption spectroscopy, with all spectra having an integration time of 1 second. It was ensured beforehand that the signal remained stable over time by monitoring it for 1 hour, confirming consistency once the jet was heated up. An initial spectrum of the broadband lamp was recorded just before igniting the plasma ($I_{0}$), followed by spectra taken every minute for 5 minutes after plasma ignition ($I_{\rm T}$) ensuring the measurement was performed with the discharge in an equilibrium state. After each measurement, a minimum of 5 minutes was allowed for the absorption signal to reset to an undetectable level. The ozone density was calculated using Lambert-Beer's law using the following formula and averaged over the 5 spectra within the measurement time of 5\,min,

\begin{equation}
    n_{\rm O_{3}} = \frac{p}{k_{\rm B} \cdot T} \cdot\frac{\log(\frac{I_{0}}{I_{\rm T}})}{k \cdot d}
\end{equation}

\noindent
where $p$ is the ambient pressure, $T$ the effluent temperature in the jet (approx. 300\,K), $I_{0}$ the initial intensity of the broad band lamp passing through the filter, $I_{\rm T}$ the transmitted intensity through the gas flow cell, $k$ the absorption coefficient of ozone ($k = 333$\,cm$^{-1}$)\cite{edward-absorption-coefficient_1953} and $d$ the length of the gas flow cell ($d = 79$\,mm).

\subsection{Emission spectroscopy of {\rm O}$_{2}(\text{\rm a}^{1}\Delta_{\text{g}})$}
Figure \ref{fig:gas flow cell} shows the optical setup for the O$_{2}(\text{a}^{1}\Delta_{\text{g}})$ detection, although the UV lamp collimator has no use here. The emission of the O$_{2}(\text{a}^{1}\Delta_{\text{g}})$ at 1270\,nm was collected by the detection collimator and guided through an optical fiber into an infrared sensitive spectrometer (Avantes AVASPEC-NIR512-1.7HSC-EVO, NIR150-1.2 from 900\,nm to 1695\,nm; wavelength resolution 0,3 nm). The collimator had a diameter of approx. 10\,mm which allowed to gather the emission passing through the plane quartz windows of the gas flow cell. To calculate the O$_{2}(\text{a}^{1}\Delta_{\text{g}})$ density the spectrometer had to be calibrated beforehand. To this end, a laser diode (RLT1270-20MGS-B; PW\,=\,1270\,nm, FWHM\,=\,0.3\,nm) was used for which the intensity was measured and compared to the radiative power given by an absolutely calibrated InGaAs-photodiode (Thorlabs FGA21-CAL; wavelength range: 800\,-\,1700\,nm). First the intensity was measured using the spectrometer. Keeping the optical setup the same, the fiber was then coupled into the calibrated photodiode. This allowed us to calculate the energy-intensity calibration constant $C = \frac{I_{0}}{P \cdot t_{\rm Int}}$, where $I_{0}$ is the peak intensity of the laser diode at 1270\,nm, $P$ the measured power of the calibrated photodiode and $t_{\rm Int}$ the integration time of the spectrometer. \\

\noindent
The O$_{2}(\text{a}^{1}\Delta_{\text{g}})$ emission spectra were time integrated over 10\,s. To calculate the O$_{2}(\text{a}^{1}\Delta_{\text{g}})$ density the following equation was used,

\begin{equation}
   n_{{\rm O}_{2}({\rm a}^{1}\Delta_{g})} = I \cdot C \cdot \frac{1}{A_{ik} \cdot Q} \cdot \frac{\lambda}{h \cdot c} \cdot \frac{1}{V} \cdot \frac{1}{f(\mathrm{d}\Omega) \cdot g}
\end{equation}

\noindent
where $I$ is the intensity, $C$ is the wavelength dependent energy-intensity calibration constant of the spectrometer, $V$ the volume of the flow cell. $f(\mathrm{d}\Omega)$ the geometric factor of how many photons produced inside the gas flow cell are reaching the detector and $g$ the loss factor of the optical components. $Q = \frac{A_{ik}}{A_{ik} + q}$ is the photon yield, which relates the observable radiative transition to the non-radiative decay of the O$_{2}(\text{a}^{1}\Delta_{\text{g}})$ molecule. $q = \sum\limits_{i} k_{i} n_{i}$ is the quenching rate. \\

\begin{table*}[t]
    \centering
    \caption{Quenching coefficients of different reactive species with O$_{2}(\text{a}^{1}\Delta_{\text{g}})$ calculated using their respective densities and rate coefficients. Rate coefficients are taken from \cite{stafford_o_2004}.}
    \begin{tabular}{l|c|c|c}
      Reaction  & $q_{i}$\,/\,s$^{-1}$ & $k_{i}$\,/\,cm$^{3}$\,s$^{-1}$ & $n_{i}$\,/\,cm$^{-3}$ \\
      \hline
       O$_{2}(a^{1}\Delta_{g})$ + O$_{2}$ $\longrightarrow$ 2\,O$_{2}$  & 0.1 & $1.5 \times 10^{-18}$ & $5.0 \times 10^{16}$ \\
       O$_{2}(a^{1}\Delta_{g})$ + O  $\longrightarrow$ O$_{2}$ + O & 0.8 & $2.0 \times 10^{-16}$ & $4.0 \times 10^{15}$ \\
       O$_{2}(a^{1}\Delta_{g})$ + O$_{3}$  $\longrightarrow$ O + 2\,O$_{2}$ & 12.1 &  $4.0 \times 10^{-15}$ & $3.0 \times 10^{15}$ \\
       O$_{2}(a^{1}\Delta_{g})$ + He  $\longrightarrow$ O$_{2}$ + He & 0.1 & $1.0 \times 10^{-21}$ & $1.0 \times 10^{19}$ \\
\end{tabular}
    \label{tab:rate_coefficients}
\end{table*}

\noindent
Table \ref{tab:rate_coefficients} shows the the quenching rates of the most prominent quenching partners for O$_{2}(\text{a}^{1}\Delta_{\text{g}})$. For our case, ozone possesses by far the highest rate coefficient, which simplifies the expression  to $q = k_{\rm O_{3}} \cdot n_{\rm O_{3}}$, with $k_{\rm O_{3}}$ being the quenching rate coefficient of ozone with O$_{2}(\text{a}^{1}\Delta_{\text{g}})$ and $n_{\rm O_{3}}$ the ozone density. The previously acquired data for the ozone density of the experiment is used to calculate the absolute O$_{2}(\text{a}^{1}\Delta_{\text{g}})$ density.\\

\subsection{{\rm O}$_{2}(\text{b}^{1}\Sigma_{\text{g}}^{+})$ density measurement by emission spectroscopy}

\begin{figure}[H]
    \centering
    \includegraphics[width=\linewidth]{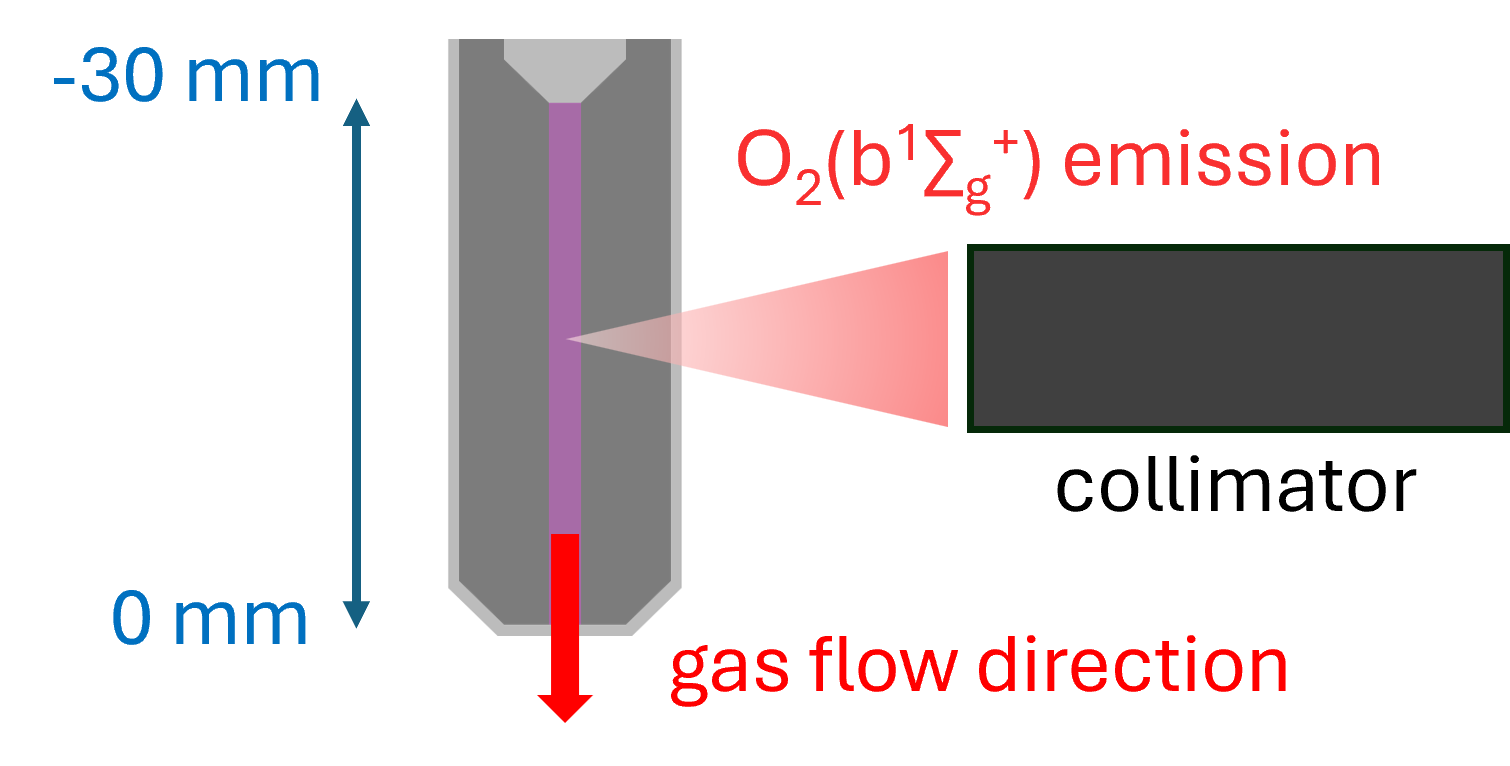}
    \caption{Emission spectroscopy setup for the O$_{2}(\text{b}^{1}\Sigma_{\text{g}}^{+})$ measurement in the jet. Red arrow indicates gas flow direction.}
    \label{fig:emission setup}
\end{figure}

The setup for the O$_{2}(\text{b}^{1}\Sigma_{\text{g}}^{+})$ emission spectroscopy is shown in figure \ref{fig:emission setup}. O$_{2}(\text{b}^{1}\Sigma_{\text{g}}^{+})$ emits at 760\,nm and was captured  with the infrared sensitive spectrometer used in the O$_{2}(\text{a}^{1}\Delta_{\text{g}})$ measurement. The emission was measured mainly in the discharge channel of the COST plasma jet because the relatively short lifetime of O$_{2}(\text{b}^{1}\Sigma_{\text{g}}^{+})$ reduces its density beyond the detection threshold quickly outside of the plasma jet. To measure the O$_{2}(\text{b}^{1}\Sigma_{\text{g}}^{+})$ density the spectrometer had to be energy-intensity calibrated similarly as described in the O$_{2}(\text{a}^{1}\Delta_{\text{g}})$ section. Here, an LED (Thorlabs LED760L; PW\,=\,760\,nm, FWHM\,=\,24\,nm) was employed, instead of a laser diode. An optical filter (CW\,=\,760\,nm, FWHM\,=\,10\,nm) was placed between diode and spectrometer, to get a more comparable power measurement between calibrated diode and spectrometer, as the width of the O$_{2}(\text{b}^{1}\Sigma_{\text{g}}^{+})$ spectrum is similar to the spectral width of the filter. The intensity of the spectrometer signal was then integrated over the spectral width and compared to the energy of a calibrated Si-photodiode (Thorlabs FDS100; wavelength range: 350\,-\,1100\,nm). The optical setup was unchanged when switching the fiber between spectrometer for the intensity measurement and the photodiode for the power measurement.

\section{Computational methods}
The two simulations presented use a very similar chemical kinetics scheme, leading to a similar prediction about the species distribution inside the plasma chamber. Only simplified wall reactions were taken into account for the reactive species or, in the case of ozone, were completely neglected. The used wall reaction rates were optimized for the plasma region of the COST jet. Investigation of the correct wall rate coefficient for various species was beyond the scope of this study.

\subsection{Pseudo-1D plug flow model}
A pseudo-one-dimensional (pseudo-1D) plug flow model is used in this study to calculate the spatially resolved species' densities along the gas flow direction in the jet chamber \cite{he_atomic_2025} \cite{he_zero_2021}. The particle balance equations and electron energy balance equation of the zero-dimensional simulation are solved to output the time-resolved species densities and an effective electron temperature in an infinitesimal plug volume. This plug volume co-moves with the gas flow, which allows the temporal variation to be converted into a spatial variation through knowledge of the gas flow velocity. \\

The gas temperature in the jet chamber is varied as a function of the absorbed power in the simulations, using equation (1) in \cite{he_atomic_2025}. The gas temperature of the feed input gas and that in both effluent regions (i.e. 17.5\,mm long tube and 79\,mm long gas flow cell in Figure \ref{fig:gas flow cell}) are approximated to be 300\,K. The pressure is assumed to be $1\times10^5$\,Pa in all regions. \\

The plug flow model calculations in both effluent regions are carried out in a similar way to those in the jet chamber. However, the model is modified due to the different geometry and power between the chamber and effluent. In particular, the cross section area of the rectangular jet chamber ($1\times1$\,mm$^2$) is smaller than those of both cylindrical effluent regions ($1\times1 \: \pi$\,mm$^2$ and $7\times7 \: \pi$\,mm$^2$). Because of this, the flow velocity in both effluent regions is calculated by replacing the gas temperature, pressure and cross section area of the jet chamber in equation (5) in \cite{he_zero_2021}, with those of the corresponding effluent region in the current study. Therefore, the gas flow velocity in the jet chamber is larger than those in both effluent regions, e.g. at 1\,slm He diluted with $0.5\,\%$ O$_2$, the flow velocities in the jet chamber, the 17.5\,mm long entry tube and 79\,mm long gas flow cell are calculated to be around 1882\,cm\,s$^{-1}$, 533\,cm\,s$^{-1}1$ and 11\,cm\,s$^{-1}$, respectively. The flow velocity value used in plug flow model in the jet chamber is similar to that calculated by the fluid model presented in Figure \ref{fig:mesh}, while the velocities of the plug flow model in both effluent regions underestimate those calculated by the fluid model. The underestimation in the 79\,mm long gas flow cell is likely ascribed to the fact that the gas flow direction in the gas flow cell region assumed in the plug flow model is the same as those in the jet chamber and tube (i.e. it is not perpendicular to those in the jet chamber and tube, as illustrated by the experimental setup shown in Figure \ref{fig:gas flow cell}). This is less realistic compared to the gas flow direction simulated by the fluid model. \\

The reaction rate of a neutral species at the wall is calculated by replacing the effective diffusion length \cite{chantry_a_1987}, the volume and the net surface area with those for the corresponding effluent region of the gas flow cell setup. Furthermore, the electron energy balance equation (i.e. equation (9) in \cite{he_zero_2021}) is excluded in the simulations of both effluent regions due to the absence of the power and the resulting negligible effect of the electron density. Therefore, only the particle balance equations are solved in the simulations of both effluent regions. \\

A Boltzmann solver based on the steady-state solution of the Boltzmann equation under the two-term approximation, i.e. the open-source simulation tool Lisbon Kinetics Boltzmann (LoKI-B) published by Tejero-del-Caz \textit{et al.} \cite{tejero-del-Caz_the_2019}, is used to calculate the non-Maxwellian electron energy distribution functions (EEDFs) for a range of values of the applied reduced electric field. \\

The plasma-chemical kinetics scheme considered in this work is identical to that used in \cite{he_atomic_2025}, where the lists of all the species and the reactions are provided. Specifically, $\mathrm{He}$, $\mathrm{He(2 ^3S)}$, $\mathrm{He_2^*}$, $\mathrm{He^+}$, $\mathrm{He_2^+}$, $\mathrm{O(^3 P)} $, $\mathrm{O_2}$($v=0$), $\mathrm{O_3}$, $\mathrm{O(^1 D)}$, $\mathrm{O_2 (a^1 \Delta_g)}$, $\mathrm{O^+}$, $\mathrm{O_2^+}$, $\mathrm{O_4^+}$, $\mathrm{O^-}$, $\mathrm{O_2^-}$, $\mathrm{O_3^-}$, $\mathrm{O_4^-}$, e$^-$, and $\mathrm{O_2 (b^1 \Sigma_{g^+})}$ are included. It is worth to note that the simulated $\mathrm{O(^3 P)}$ densities produced in He/O$_2$ plasma jets in \cite{he_atomic_2025} are well validated against measured densities from multiple publications under variations of the absorbed power, gas flow rate, and mixture ratio. \\

\subsection{Fluid model}
The computational model employed in this study to obtain spatially resolved data in two dimensions is {\it nonPDPSIM}, a fluid-based plasma simulation code \cite{kushner_modeling_2004,kushner_modelling_2005}. The code solves the continuity equation for charged and neutral species assuming the drift-diffusion approximation and the compressible Navier-Stokes equations providing the species densities, flow field and the gas temperature distribution, on an unstructured triangular grid. The charged species considered in the model are electrons (e$^-$), positive (O$_{2}^{+}$, O$^{+}$, He$^{+}$), and negative ions (O$^{-}$, O$_{2}^{-}$, O$_{3}^{-}$), while neutral species include He, O$_2$, O, O$_3$, O$_2$($v$=1-4) (first four vibrational levels of O$_2$), O$_3$($v$), O$_2$(a$^{1}\Delta_{\mathrm{g}}$), O$_2$(b$^{1}\Sigma_{\mathrm{g}}^{+}$), O($^1$D), and He$^\ast$ (which is an ensemble of He(2$^3$S) and He(2$^1$S)). The transport coefficients for heavy species (i.e., not electrons) are calculated by assuming that the respective species temperature is equal to the gas temperature. The electron transport coefficients (mobility and diffusion) are calculated from a 2-term Boltzmann solver. The simulation uses the local mean energy approximation and therefore solves the energy equation as well. The chemistry set and the cross sections for the Boltzmann solver are identical to that reported in \cite{liu_local_2023}. \\

\begin{figure}[H]
    \centering
    \includegraphics[width=\linewidth]{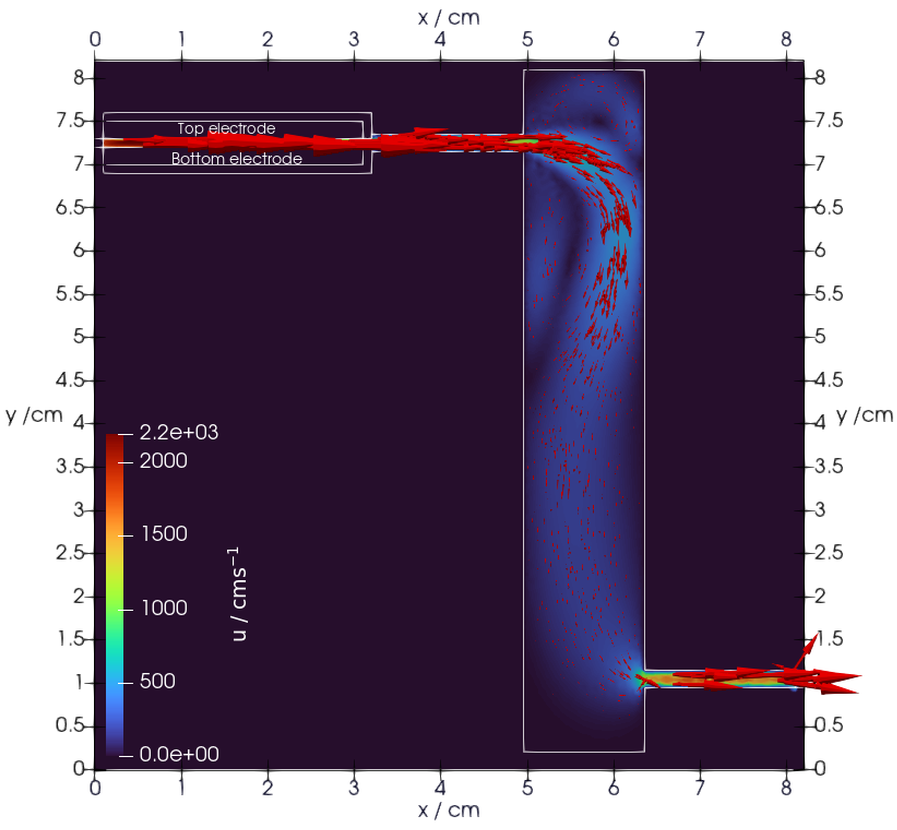}
    \caption{Computational domain and spatial distribution for the gas velocity at 1\,slm He flow. The color bar shows the gas velocity while the red arrows indicate the gas flow direction.}
    \label{fig:mesh}
\end{figure}

Fig. \ref{fig:mesh} illustrates the computational domain along with the spatial distribution of the gas flow field at a He inflow of 1\,slm after the stationary state has already been attained. The neutral species generated within the jet (with the ``Top electrode'' driven and the ``Bottom electrode'' grounded) flow through a tube with a 2\,mm gap length, before entering the flow cell, as shown in the figure. The dimensions of the computational domain agree with the experimental setup described above. As seen in the Figure, the gas velocity distribution is non-uniform in the flow cell. Note that, due to the change in gap length between the jet electrodes and the tube leading to the flow cell, the velocity decreases by a factor of 2 (as the simulation is two-dimensional). Moreover, as the gas enters the flow cell, its velocity further decreases, forming two vortices near the inlet. \\

\section{Results}

\subsection{Ozone density}
\begin{figure}
\centering
\includegraphics[width=.5\textwidth]{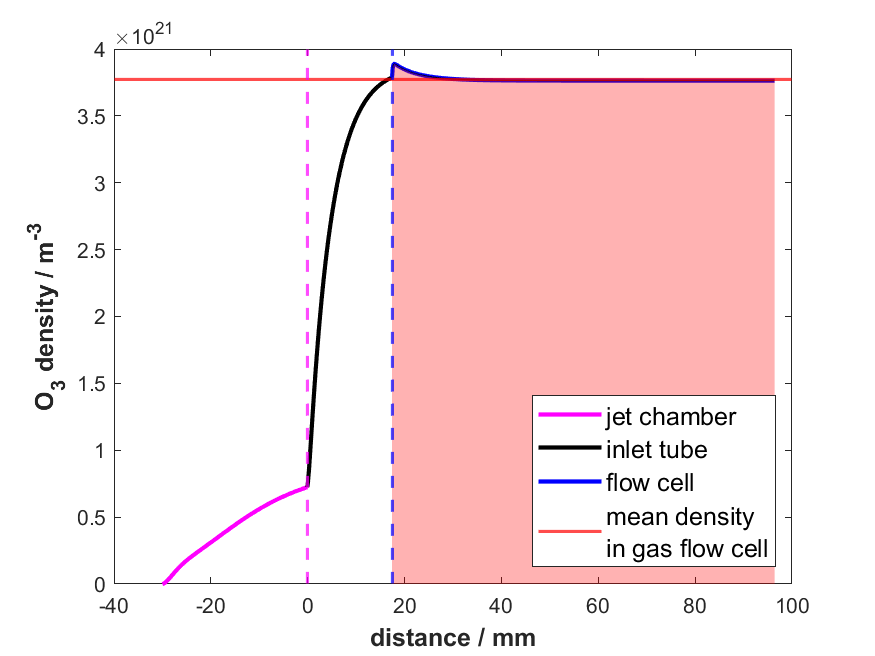}
\caption{O$_{3}$ density calculated by the plug-flow model along the gas flow direction in and outside the plasma jet. Averaged density in the gas flow cell volume is shown as a red line. Dashed lines show the transition from the jet to the inlet tube (pink) and to the flow cell (blue). An effluent temperature of 300\,K, a power of 1\,W, a 1\,slm He flow and a 0.5\,\% oxygen admixture are assumed.\label{fig:plug-flow_integ_ozone}}
\end{figure}

\begin{figure}
\centering
\includegraphics[width=.5\textwidth]{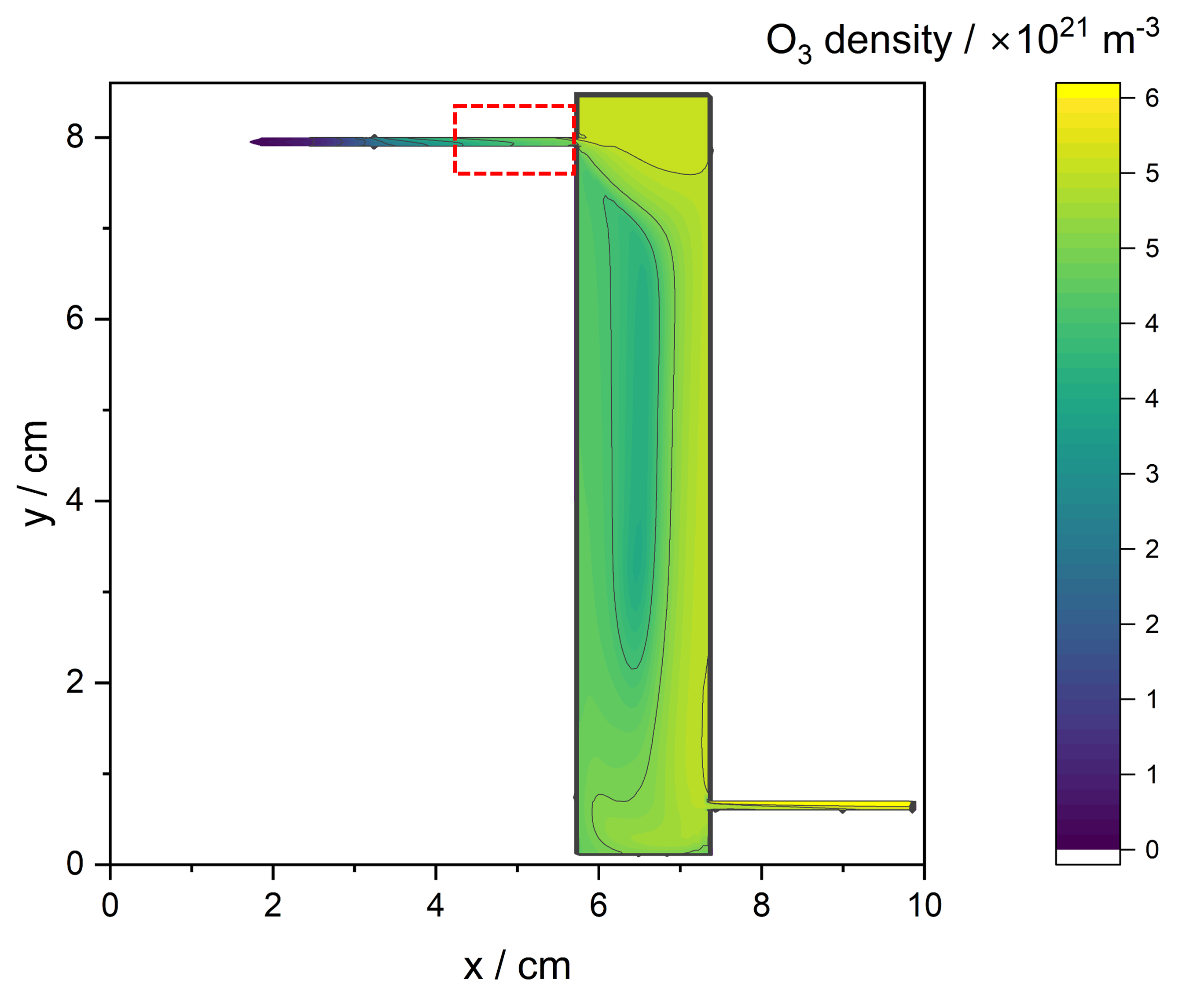}
\caption{O$_{3}$ density distribution inside the gas flow cell simulated by a fluid model for a effluent temperature of 300\,K, power of 1\,W, flow of 1\,slm He and a 0.5\,\% oxygen admixture. The red dotted area marks the inlet tube after the plasma jet channel.\label{fig:2D O3 1W}}
\end{figure}

The ozone density was studied under varying power, flow rate, and O$_{2}$ admixture conditions. Experimental results were compared with simulations from both the pseudo-1D plug flow model and the 2D fluid model. The plug flow model focuses on plasma chemistry within the jet and effluent, as illustrated in Figure \ref{fig:plug-flow_integ_ozone}, while the 2D fluid model accounts for both chemistry and gas dynamics in the flow cell geometry, as shown in Figure \ref{fig:2D O3 1W}.\\

\noindent
Figure \ref{fig:plug-flow_integ_ozone} shows the ozone density calculated by the plug flow model along the gas flow direction, where $x = 0$\,m is the end of the jet electrodes. In the plasma channel, the ozone density increases asymptotically, reaching a maximum at the jet nozzle. Along the length of the jet, oxygen is dissociated into atomic oxygen. The atomic oxygen then reacts with molecular oxygen to form ozone, which is one of the most prevalent production mechanisms. With the increased dissociation degree along the jet more atomic oxygen is available to produce ozone \cite{steuer_2d_2021}. The production would eventually equalize with the reaction of ozone with O$_{2}(\text{b}^{1}\Sigma_{\text{g}}^{+})$. However, for a flow of 1\,slm the residence time in the plasma region is not long enough to reach an equilibrium. \\

\noindent
In the inlet tube to the gas cell, the ozone density continues to rise asymptotically. The fast rise, with respect to distance, of the ozone density from the transition of the jet chamber to the inlet tube occurs because of the slower gas velocity in the inlet tube. This leads to a strong increase in ozone density with distance as the residence time in the inlet tube is longer for the same path length. Furthermore, ozone destruction rates like reactions with O$_{2}(\text{b}^{1}\Sigma_{\text{g}}^{+})$ or electron impact dissociation are lower in the effluent region.\\

\noindent
Upon entering the flow cell the gas velocity decreases further, which leads to a spike in the density because of the significantly higher residence time. Due to the minimal influence of destruction reactions and ozone's long lifetime the density remains approximately constant. The plasma chemistry therefore converges to the more long lived species like ozone and molecular oxygen. The ozone density in the flow cell is averaged along the flow direction vector, as shown in Figure \ref{fig:plug-flow_integ_ozone}. This averaged value is then compared to the experimental results. \\

\noindent
Figure \ref{fig:2D O3 1W} shows the two dimensional ozone density distribution in the gas flow cell as simulated by the 2D fluid model. The effluent is then directed into a tube that connects vertically to the gas flow cell, as described in chapter \ref{chp:flow cell}. As shown in the plug flow simulation, the ozone density increases significantly outside the plasma zone, especially in the flow cell. In the cell, the flow follows a vortex along the cell walls, resulting in a maximum density at the outer sides of the flow cell. In the center axis of the cell, the density is lower due to the vortex effect. Additionally, the pocket at the entrance, where the optical window is located, also generates a vortex, leading to a slightly higher density in that area compared to the rest of the cell. The inhomogeneous distribution of ozone by the vortex is a result of the relatively short simulation time ($\approx 0.2$\,s). With a prolonged simulation time, the modeled density of long-lived oxygen species like ozone is expected to reach a more homogeneous distribution inside the cell. This is primarily due to diffusion: since the diffusion coefficient of ozone in helium is $D_{\rm O_3}=7.13\cdot10^{-5}$\,m$^2$/s \cite{waskoenig_atomic_2010}, the timescale for homogenization would be on the order of seconds. Unfortunately, increasing the simulation time to a sufficient amount was computationally very expensive and beyond the scope of this study. \\

\begin{figure}
\centering
\includegraphics[width=.5\textwidth]{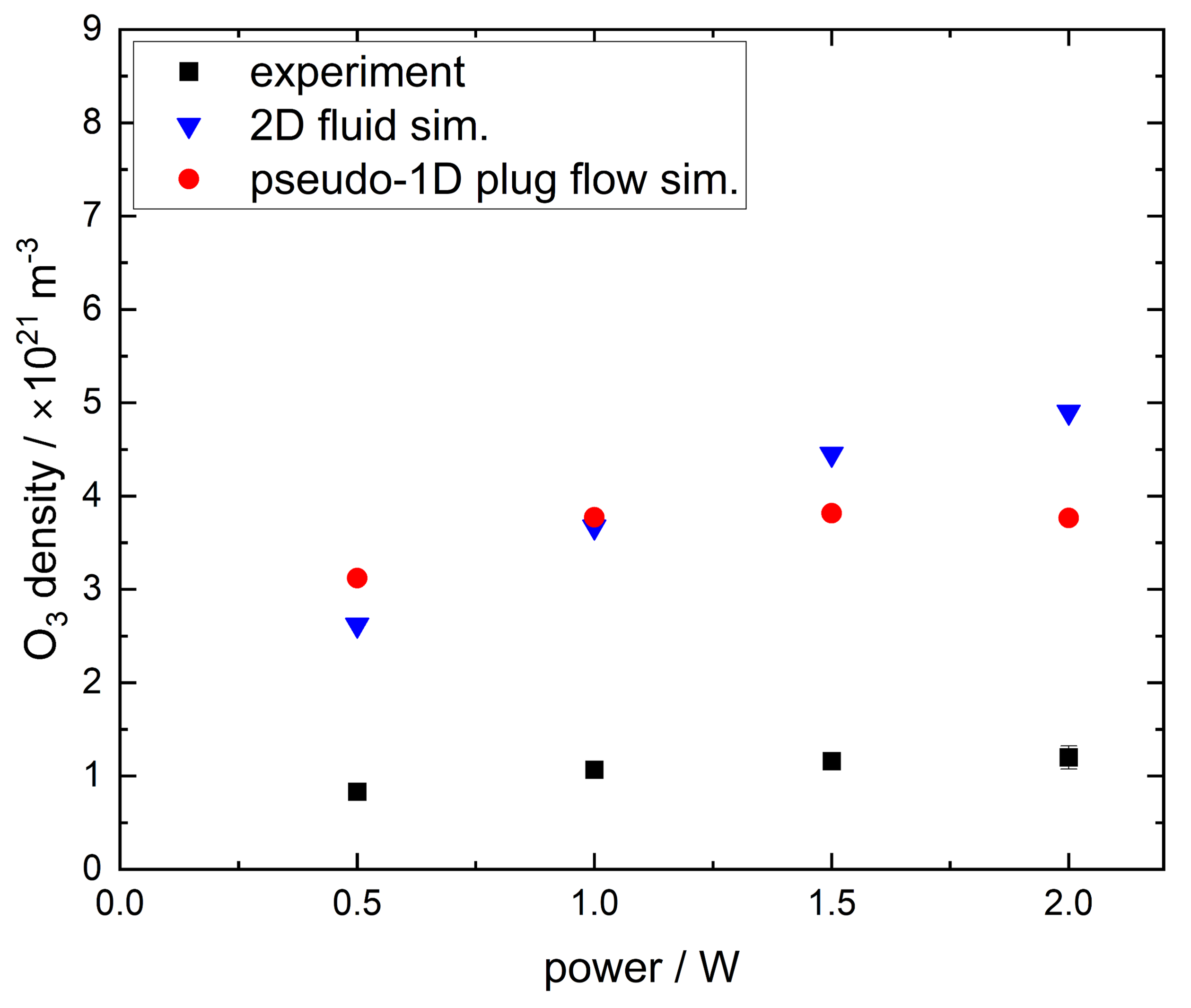}
\caption{O$_{3}$ density comparison between simulation and experiment for a effluent temperature of 300\,K, flow of 1\,slm He and a 0.5\,\% oxygen admixture depending on the plasma power.\label{fig:O3 power}}
\end{figure}

\noindent
Figure \ref{fig:O3 power} shows the ozone density as a function of plasma power for a helium flow of 1\,slm and an oxygen admixture of 0.5\,\%. Both simulations and experimental results demonstrate a gradual increase in ozone density with rising plasma power. The increased plasma power leads to a higher dissociation degree of molecular oxygen and therefore to a higher ozone density as more atomic oxygen is available. The experimental data show a linear increase of the ozone density with power. This trend is matched by the 2D fluid simulation, although its slope is higher than that observed for the experimental values. The density values of the two simulations match quite well but are approx. a factor of three higher than what the experimental data suggests, though they remain within the same order of magnitude. The plug flow simulation also shows a more asymptotic increase. The ozone density reaches a plateau at 1\,W. Here the ozone production is likely equilibrated by reactions with O$_{2}(\text{b}^{1}\Sigma_{\text{g}}^{+})$), which prevents the ozone density from increasing further. The good agreement in the density values of the two simulations supports the assumption that the volume-averaged density in the plug flow simulation gives a reasonable agreement under the condition that the density is homogeneously distributed within the gas flow cell. One possible explanation for the discrepancy between the simulations and experimental results could be related to the reaction dynamics within the flow cell. For instance, the simulations do not account for ozone's wall destruction reactions. Given the similarity in the simulated ozone densities, it is plausible that both simulations underestimate the destruction rate of ozone and or overestimate its production, resulting in higher simulated values compared to the experiment. Additionally, ozone is sensitive to conditions such as temperature. Small deviations in experimental temperature from the assumed 300\,K effluent temperature in the simulations could also contribute to the lower ozone density observed experimentally \cite{wijaikhum_absolute_2017,bang_temperature-dependent_2023}.\\

\begin{figure}
\centering
\includegraphics[width=.5\textwidth]{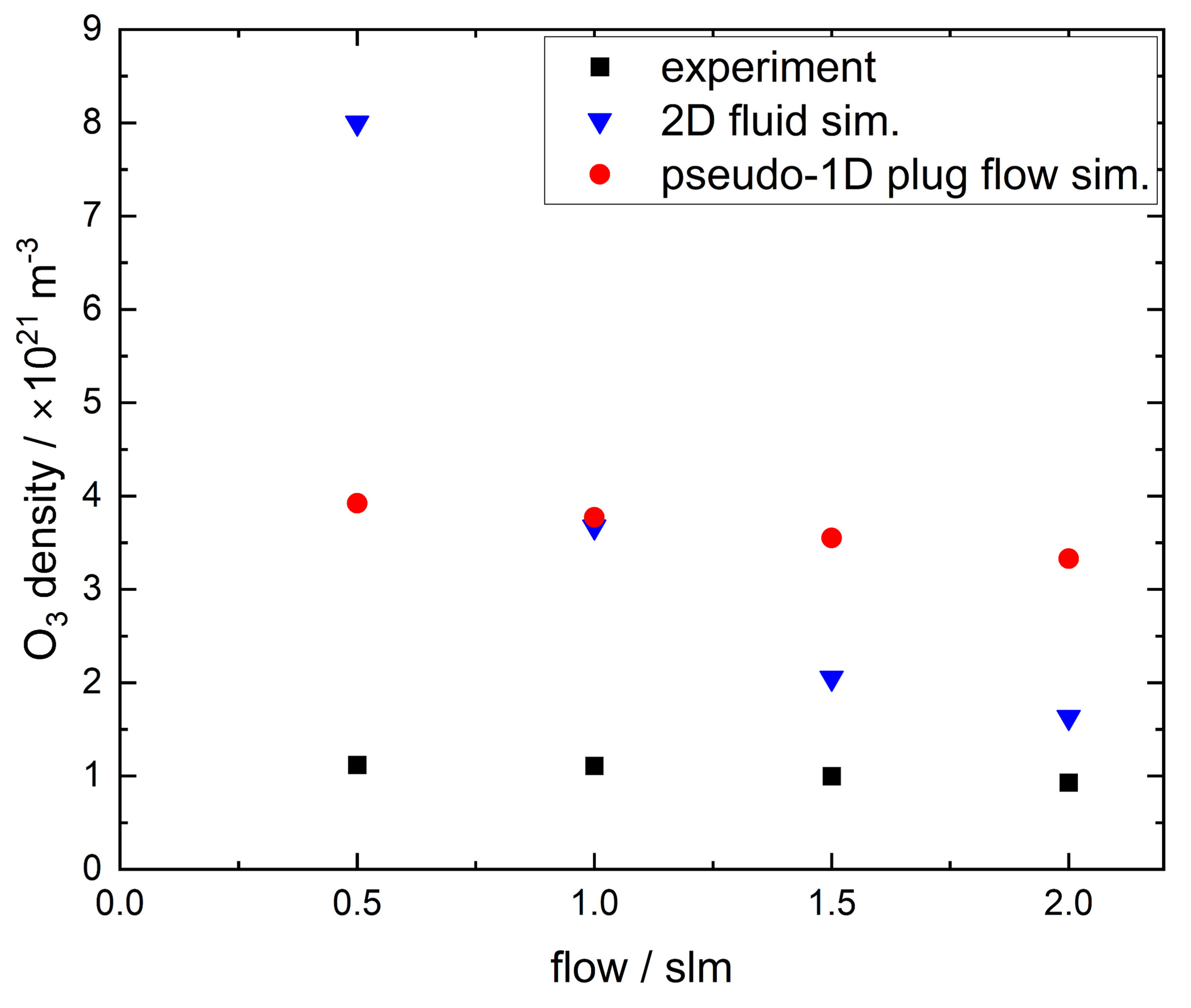}
\caption{O$_{3}$ density comparison between simulation and experiment for a effluent temperature of 300\,K, power of 1\,W plasma power and a 0.5\,\% oxygen admixture depending on the helium flow.\label{fig:O3 flow}}
\end{figure}

\noindent
The ozone density is influenced by the gas flow velocity, as shown in Figure \ref{fig:O3 flow}. Both experimental results and simulations agree that the ozone density is highest at low flow rates and decreases as gas velocity increases. This is due to the longer residence time in the plasma volume at lower flows. Ozone is primarily formed through the reaction of atomic oxygen with molecular oxygen, in the presence of a third body. With a lower gas flow, the longer residence time in the plasma region leads to a higher dissociation degree of oxygen, increasing the atomic oxygen density and, consequently, the ozone density. Both the experiment as well as the plug flow simulation show a linear decrease with the flow velocity with an almost identical slope. The 2D fluid simulation however predicts a more asymptotic decrease. An explanation might be that the fluid simulation does not reach a homogeneous distribution in the gas flow cell within the simulation time as shown in Figure \ref{fig:2D O3 1W}. This explains the high discrepancy for low flows, as here the time to reach an equilibrium is much higher than for higher flows. Higher flows on the other hand should reach the equilibrium faster and indeed have a smaller deviation from experimental data. As observed previously in the power variation measurements, the simulations still overestimate the ozone density compared to the experiment. It should however be noted that the 2D fluid simulation density values come very close to the experimental values for high flows, as the ozone distribution gets more homogeneous. With longer computation time the 2D fluid simulation might therefore be able to provide an accurate description of the experimentally measured densities. \\

\begin{figure}
\centering
\includegraphics[width=.5\textwidth]{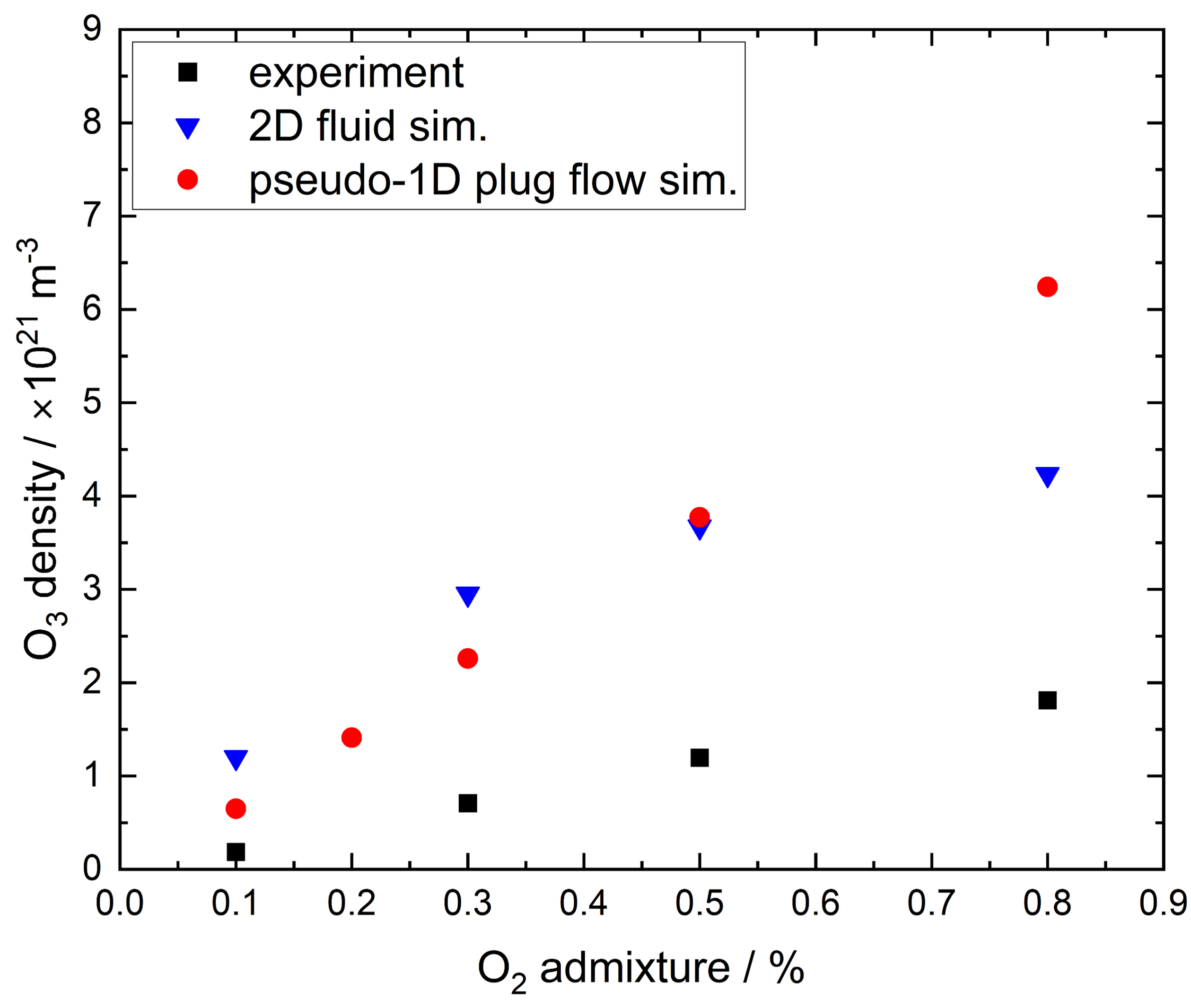}
\caption{O$_{3}$ density comparison between simulation and experiment for a effluent temperature of 300\,K, power of 1\,W plasma power and a helium flow of 1\,slm depending on the oxygen admixture.\label{fig:O3 O2}}
\end{figure}

\noindent
Figure \ref{fig:O3 O2} shows the ozone density as a function of oxygen admixture. Both simulation and experimental results demonstrate an ozone density increase with the oxygen admixture. Ozone is a product of the recombination of atomic oxygen with molecular oxygen. It is well known that the COST jet has an optimal oxygen admixture for atomic oxygen production around 0.5\,\% in helium \cite{steuer_2d_2021}. Lower admixtures result in a lower atomic oxygen density because less molecular oxygen can be dissociated, while higher admixtures reduce the atomic oxygen density because the dissociation of molecular oxygen is less efficient and recombination to ozone occurs more rapidly. Even when the atomic oxygen density decreases with higher oxygen admixtures, the ozone density still rises slightly for higher admixtures, as atomic oxygen has more molecular oxygen it can react with. Both simulations predict a rise in the ozone density which agrees with the experimental data. Furthermore the ozone density values predicted by the two simulations fit quite well with the experimental results for low admixtures. High admixtures however lead to an overestimation of the ozone density. The 2D fluid simulation provides a more accurate description of the dependency on the admixture observed in the experimental results with a more asymptotic increase. The plug flow simulation however predicts an almost linear increase in the ozone density, which especially for high flows leads to a discrepancy to the experimental values. This indicates that the species distribution induced by the gas dynamics may play a more an important role for high oxygen admixtures as low admixtures generally have a better agreement with the experimental values. \\

\noindent
Both simulations provide a good estimate of the measured densities within the same order of magnitude, although overestimating the ozone density by a factor of around three. Still, the 2D fluid simulation is able to reproduce the more complex dependencies of the ozone density on the admixture variation and, given more time to converge, maybe also for the flow variation, as higher flows have shown a better agreement with experimental results. The plug flow simulation was able to replicate the trends of the power variation and flow variations but represents the trend in the admixture variation less accurately.

\subsection{\rm O$_{2}(\text{a}^{1}\Delta_{\text{g}})$ density}
The experimentally determined O$_{2}(\text{a}^{1}\Delta_{\text{g}})$ density was compared to the plug flow simulation and the 2D fluid simulation as with the ozone density. \\

\begin{figure}
\centering
\includegraphics[width=.5\textwidth]{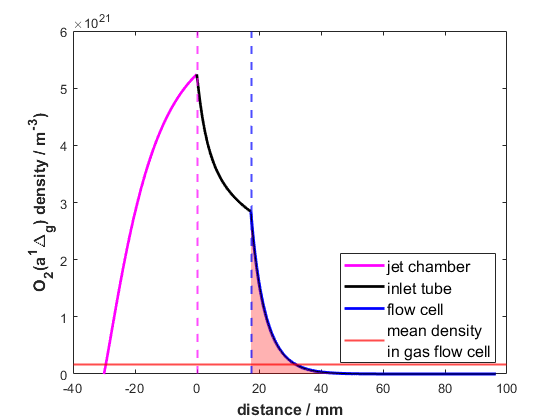}
\caption{O$_{2}(\text{a}^{1}\Delta_{\text{g}})$ density calculated by the plug flow model along the jet and effluent in the gas flow cell. Dashed lines show the transition from the jet to the inlet tube (pink) and to the flow cell (blue). Averaged density for the gas flow cell volume is shown as a red line. An effluent temperature of 300\,K, power of 1\,W, 1\,slm He flow and a 0.5\,\% oxygen admixture are assumed.\label{fig:plug-flow_integ} }
\end{figure}

\noindent
The simulated O$_{2}(\text{a}^{1}\Delta_{\text{g}})$ density for the plug flow model is shown in figure \ref{fig:plug-flow_integ}. It can be observed that within the discharge the O$_{2}(\text{a}^{1}\Delta_{\text{g}})$ density rises asymptotically towards the end of the jet. From there the gas flow transitions into the wider inlet tube ($d=2$\,mm) to the flow cell as indicated by the left dashed line. The flow cell begins after the second dashed line. In the effluent, the O$_{2}(\text{a}^{1}\Delta_{\text{g}})$ decays exponentially as it reacts. This behavior can be observed for multiple species like atomic oxygen \cite{steuer_2d_2021}. The decay rate, with respect to distance, increases when entering the gas flow cell. The bigger cross section in th flow cell results in a slower gas velocity so the O$_{2}(\text{a}^{1}\Delta_{\text{g}})$ decays faster over the same length scale. \\

\noindent
To compare the plug flow model density to the experimental values, the O$_{2}(\text{a}^{1}\Delta_{\text{g}})$ density was averaged along the gas flow direction after the inlet tube, as shown by the red area in Figure \ref{fig:plug-flow_integ}. The O$_{2}(\text{a}^{1}\Delta_{\text{g}})$ density is not homogeneously distributed inside the flow cell, but later comparison will show that this assumption still leads to a reasonably good agreement between plug flow simulation and experimental results. 
\\

\begin{figure}
\centering
\includegraphics[width=.5\textwidth]{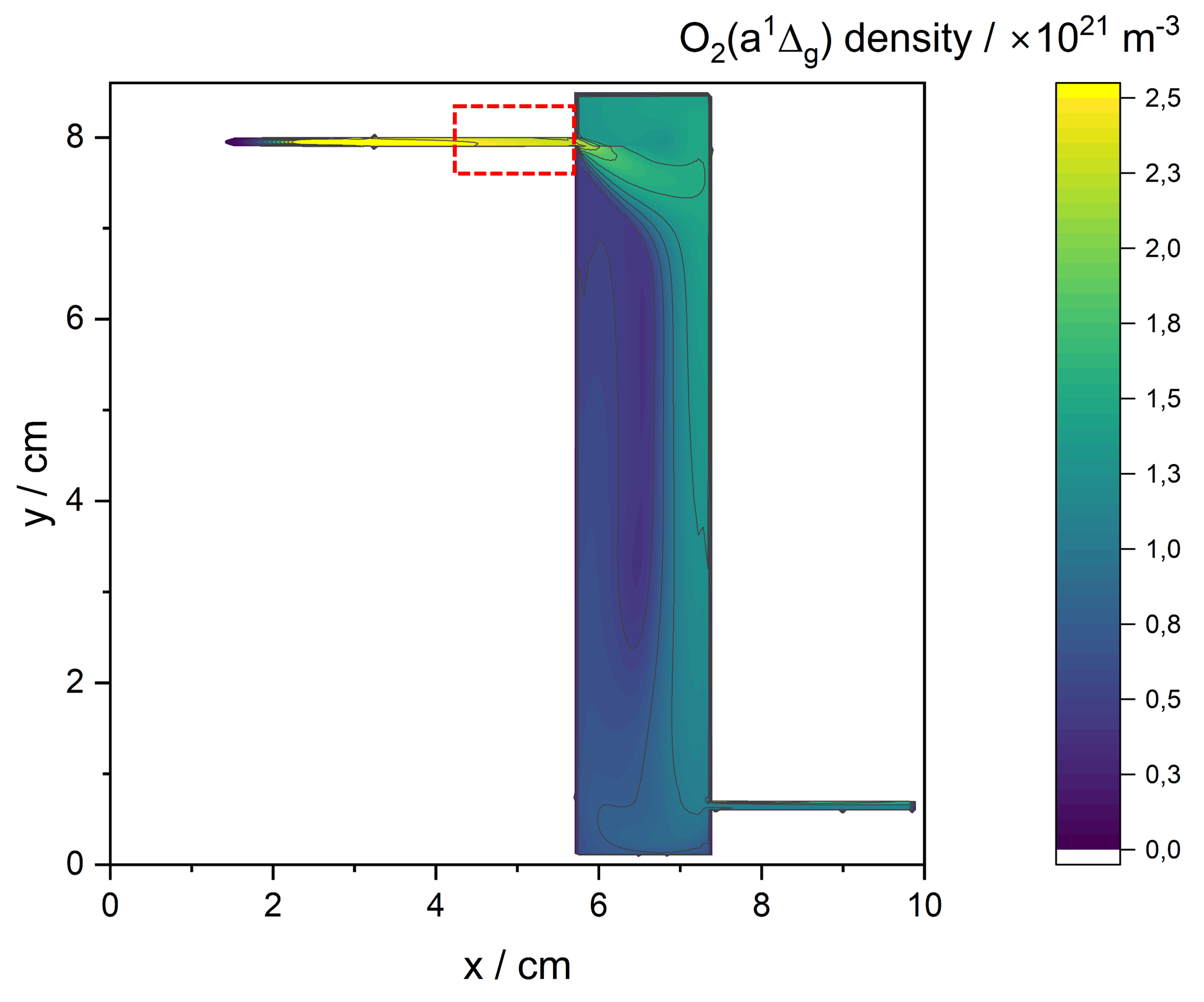}
\caption{Two-dimensional O$_{2}(\text{a}^{1}\Delta_{\text{g}})$ density in the gas flow cell volume calculated by the 2D fluid simulation of the COST plasma jet for effluent temperature of 300\,K, power of 1\,W, 1\,slm He flow and a 0.5\,\% oxygen admixture.\label{fig:2D SDO 1W}}
\end{figure}

\noindent
Figure \ref{fig:2D SDO 1W} shows the density distribution of the O$_{2}(\text{a}^{1}\Delta_{\text{g}})$ inside the gas flow cell geometry. The O$_{2}(\text{a}^{1}\Delta_{\text{g}})$ decays along the entrance tube to the flow cell outside the plasma before it reaches the cell. As the lifetime of O$_{2}(\text{a}^{1}\Delta_{\text{g}})$ is shorter than that of ozone and it reacts at a high rate with ozone, the distribution of O$_{2}(\text{a}^{1}\Delta_{\text{g}})$ is more inhomogeneous. Its density also generally decreases with increasing distance from the inlet. At the point of entering the flow cell the O$_{2}(\text{a}^{1}\Delta_{\text{g}})$ density is around $2.4\,\times\,10^{21}\,~\,\text{m}^{-3}$, which is in good agreement with the plug flow simulation values shown in Figure \ref{fig:plug-flow_integ} where the density is around $3.0 \times 10^{21} ~ \text{m}^{-3}$. At the point of leaving the flow cell, the density is almost halved to around $1.3\,\times\,10^{21}\,~\,\text{m}^{-3}$ which is much higher than the plug flow simulation predicts because of the slow gas velocity. As observed for ozone the O$_{2}(\text{a}^{1}\Delta_{\text{g}})$ density along the flow cell wall opposite to the entrance tube is slightly higher than on the entrance side. The O$_{2}(\text{a}^{1}\Delta_{\text{g}})$ therefore follows the vortex with a decrease of the density in the middle of the cell. To compare the results to the experiment the density was averaged over the volume of the gas flow cell as described for the determination of the ozone density. \\

\noindent
To study the O$_{2}(\text{a}^{1}\Delta_{\text{g}})$ density produced by the COST plasma jet we conducted variations of the plasma power, gas flow, and oxygen admixture. \\

\begin{figure}
\centering
\includegraphics[width=.5\textwidth]{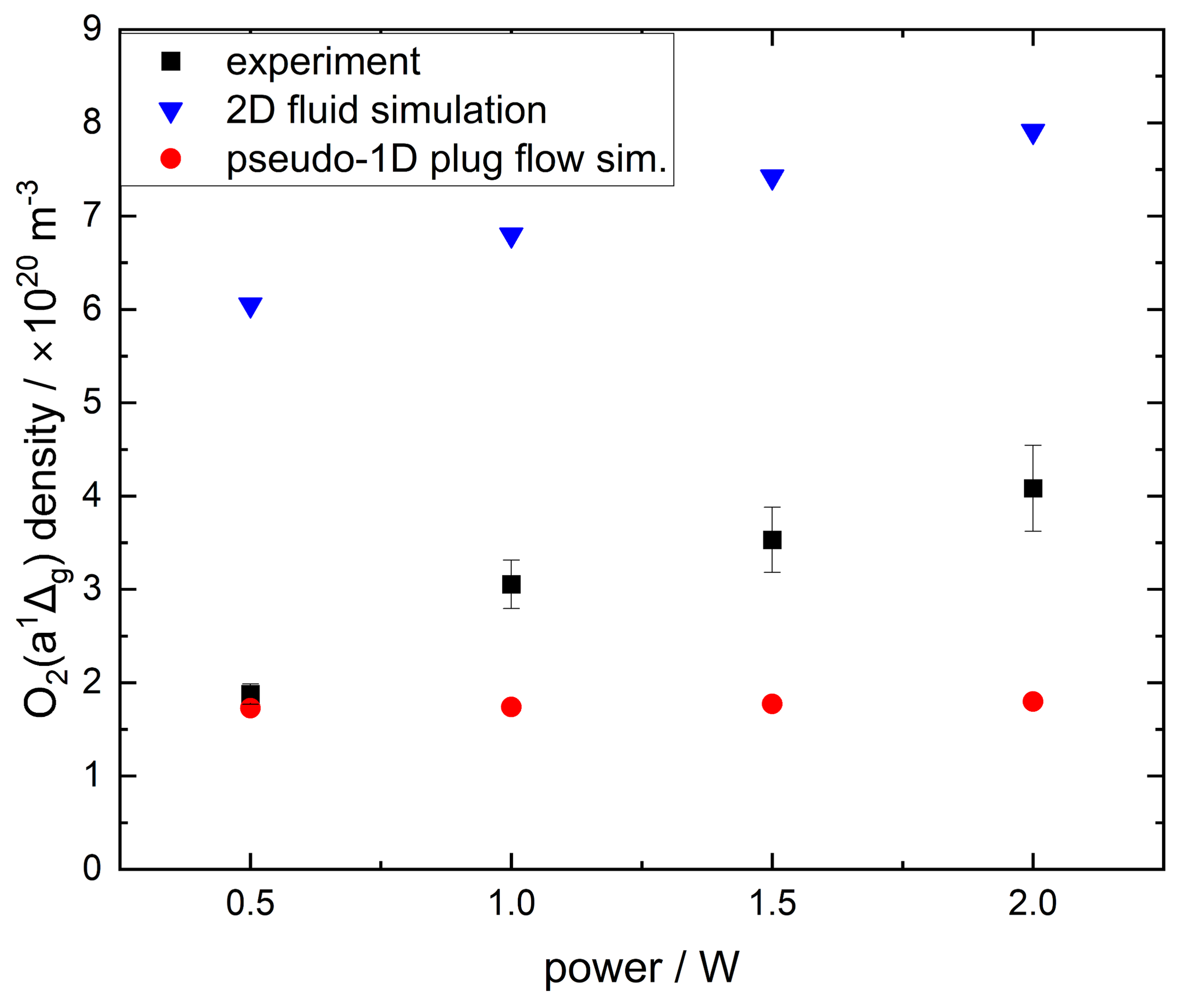}
\caption{O$_{2}(\text{a}^{1}\Delta_{\text{g}})$ density comparison between simulation and experiment for an effluent temperature of 300\,K, power of 1\,slm He flow and a 0.5\,\% oxygen admixture as a function of the plasma power. \label{fig:SDO power}}
\end{figure}

\noindent
Figure \ref{fig:SDO power} shows the O$_{2}(\text{a}^{1}\Delta_{\text{g}})$ density depending on the plasma power from experiment and simulation. The 2D fluid simulation and the experiment show an increase in the density with power. As O$_{2}(\text{a}^{1}\Delta_{\text{g}})$ is produced mostly through electron impact excitation and the reaction of O$_{2}(\text{b}^{1}\Sigma_{g}^{+})$ with ozone, it is reasonable that the density increases with higher plasma power since this also increases the ozone and O$_{2}(\text{b}^{1}\Sigma_{\text{g}}^{+})$ densities. \\

\noindent
The experimental data shows a steady increase of the O$_{2}(\text{a}^{1}\Delta_{\text{g}})$ density with the power. The same is observable for the 2D fluid simulation, although the values are roughly three times higher. However, the slope is matching very well. As mentioned before, the ozone distribution inside the flow cell does not reach a steady state within the short simulation time. This inhomogeneous distribution of the ozone might lead to a locally reduced reaction rate with O$_{2}(\text{a}^{1}\Delta_{\text{g}})$ and may result in the higher average density. \\

\noindent
The densities calculated from the plug flow simulation are within the same order of magnitude as those of the experiment and the 2D fluid simulation. However, they show only a small dependency on the plasma power, which mainly results from the slow gas velocity in the gas flow cell. Due to the long residence time in the cell and the inlet tube most of the O$_{2}(\text{a}^{1}\Delta_{\text{g}})$ already decayed. This leaves only the tail of the exponential distribution, which is not impacted strongly by an increase in the initial O$_{2}(\text{a}^{1}\Delta_{\text{g}})$ density produced inside the plasma channel. \\

\begin{figure}
\centering
\includegraphics[width=.5\textwidth]{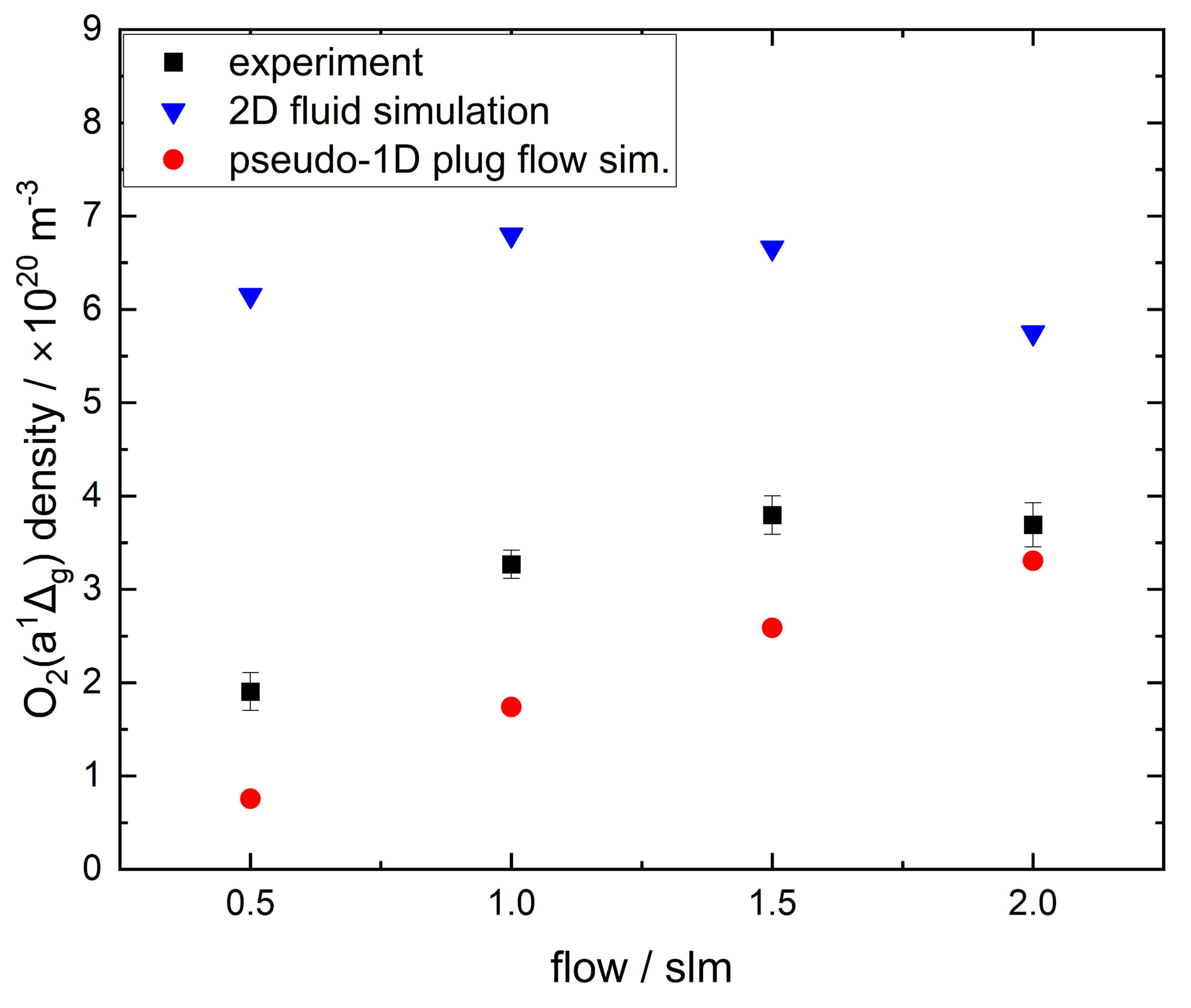}
\caption{O$_{2}(\text{a}^{1}\Delta_{\text{g}})$ density comparison between simulation and experiment for an effluent temperature of 300\,K, power of 1\,W and a 0.5\,\% oxygen admixture, depending on the gas flow.\label{fig:SDO flow}}
\end{figure}

\noindent
In Figure \ref{fig:SDO flow} the O$_{2}(\text{a}^{1}\Delta_{\text{g}})$ density is shown for a variation of the gas flow. The experimental results show an increasing O$_{2}(\text{a}^{1}\Delta_{\text{g}})$ density with the gas flow, reaching a maximum at 1.5\,slm and then slightly decreasing towards 2.0\,slm. The shorter residence time due to the increased flow velocity gives the O$_{2}(\text{a}^{1}\Delta_{\text{g}})$ less time to be quenched, thus more O$_{2}(\text{a}^{1}\Delta_{\text{g}})$ is reaching the flow cell, increasing its average density. On the other hand, with the increased gas velocity, the residence time in the plasma region decreases, which leads to less O$_{2}(\text{a}^{1}\Delta_{\text{g}})$ being formed. These two effects appear to cancel out for flow rates of 1.5\,slm to 2.0\,slm. After that, it is reasonable to assume that the density would decrease, as less and less O$_{2}(\text{a}^{1}\Delta_{\text{g}})$ can be produced in the plasma volume with the decrease in residence time. \\

\noindent
The 2D fluid simulation shows an increase in the O$_{2}(\text{a}^{1}\Delta_{\text{g}})$ density from 0.5\,slm to 1.0\,slm where it stays mostly constant before declining after 1.5\,slm. This trend can be explained similarly to the experimental results. However, in case of the fluid simulation an earlier decline in the density can be observed, which is mostly likely again due to the species distribution not reaching a complete equilibrium state within the simulation time.\\

\noindent
The plug flow simulation shows more of a linear increase in the density with its values being slightly lower than the experimental results. Because of the low gas flow velocity in the flow cell an increase in the gas flow seems to have a more pronounced influence on the averaged O$_{2}(\text{a}^{1}\Delta_{\text{g}})$ density than the decrease in production due to the shorter residence time in the plasma volume. This results in the more linear increase in the density. Flows beyond 2\,slm might still lead to decreased densities if the residence time in the plasma volume continues to decrease.\\

\begin{figure}
\centering
\includegraphics[width=.5\textwidth]{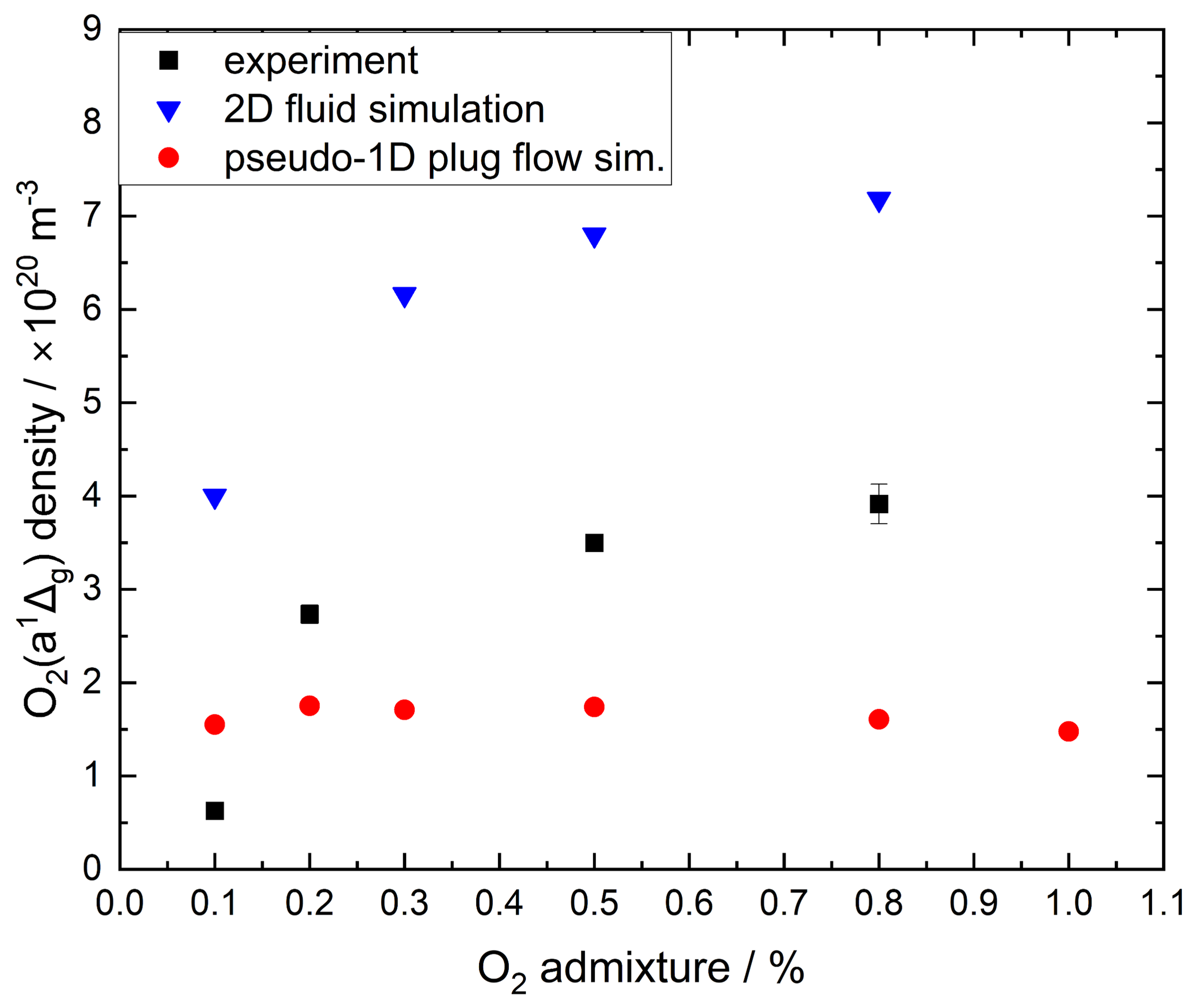}
\caption{O$_{2}(\text{a}^{1}\Delta_{\text{g}})$ density comparison between simulation and experiment for a effluent temperature of 300\,K, power of 1\,W and 1\,slm He flow depending on the oxygen admixture.\label{fig:SDO O2}}
\end{figure}

\noindent
Figure \ref{fig:SDO O2} shows the impact of the oxygen admixture on the O$_{2}(\text{a}^{1}\Delta_{\text{g}})$ density. The plasma power was kept constant at 1 W. Increasing the oxygen percentage in the gas flow leads to an asymptotic increase of the O$_{2}(\text{a}^{1}\Delta_{\text{g}})$ density in the effluent for the experimental results. This is likely due to the higher likelihood for electron impact excitation in the jet chamber if more oxygen molecules are present. The 2D fluid simulation follows the experimental trend quite closely, but has slightly higher density values.\\

\noindent
The plug flow simulation does not predict a strong dependence on the oxygen admixture. Only from 0.1\% to 0.2\% a slight increase in the density is visible, while it stays almost constant for higher admixtures. This is again more an effect of the slow gas flow velocity as the density that reaches the flow cell is quite low, so no significant change can be observed. Admixtures higher than 0.5\% even dip a little in their O$_{2}(\text{a}^{1}\Delta_{\text{g}})$ density. Here the plug flow simulation predicted higher ozone densities than 2D fluid simulation and experiment as seen in \ref{fig:O3 O2} which lead to higher destruction rates of O$_{2}(\text{a}^{1}\Delta_{\text{g}})$.\\

\noindent
The plug flow simulation shows a very good agreement with the absolute experimental values, although falls short in describing the trends correctly due to the likely underestimated gas flow velocity in the flow cell region. The 2D fluid simulation matched the experimental trends very well but overestimates the O$_{2}(\text{a}^{1}\Delta_{\text{g}})$ densities. This is attributed to the limited computation time of the simulation which does not reach an equilibrium state. The values could potentially come very close to the experiment with longer simulation time as shown for high flows which have a better agreement with the experiment.

\subsection{O$_{2}(\text{b}^{1}\Sigma_{\text{g}}^{+})$ density}
The O$_{2}(\text{b}^{1}\Sigma_{\text{g}}^{+})$ spectra were taken with an integration time of 1\,s. The O$_{2}(\text{b}^{1}\Sigma_{\text{g}}^{+})$ density in the jet was stable directly after plasma ignition if the jet had already heated up. Because the O$_{2}(\text{b}^{1}\Sigma_{\text{g}}^{+})$ is measured inside the COST jet without flow cell, no comparison to the 2D fluid simulation was conducted. \\

\begin{figure}
\centering
\includegraphics[width=.5\textwidth]{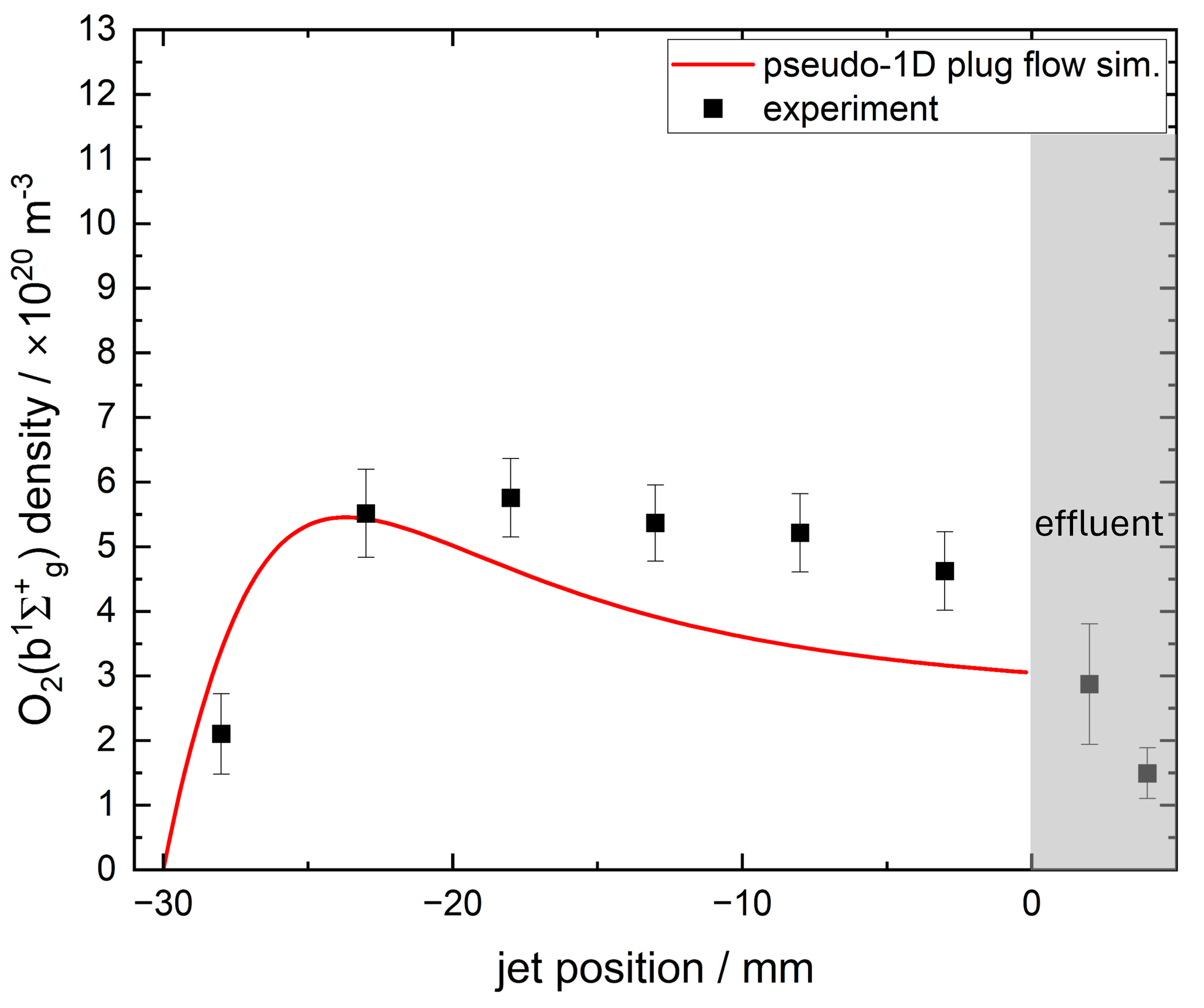}
\caption{O$_{2}(\text{b}^{1}\Sigma_{\text{g}}^{+})$ density distribution comparison between simulation and experiment. Grey area indicates effluent region. Example for a power of 1\,W , 1\,slm He flow and a 0.5\,\% oxygen admixture as a function of the discharge channel position in the jet.\label{fig:O2b pos comp}}
\end{figure}

\begin{figure}
\centering
\includegraphics[width=.5\textwidth]{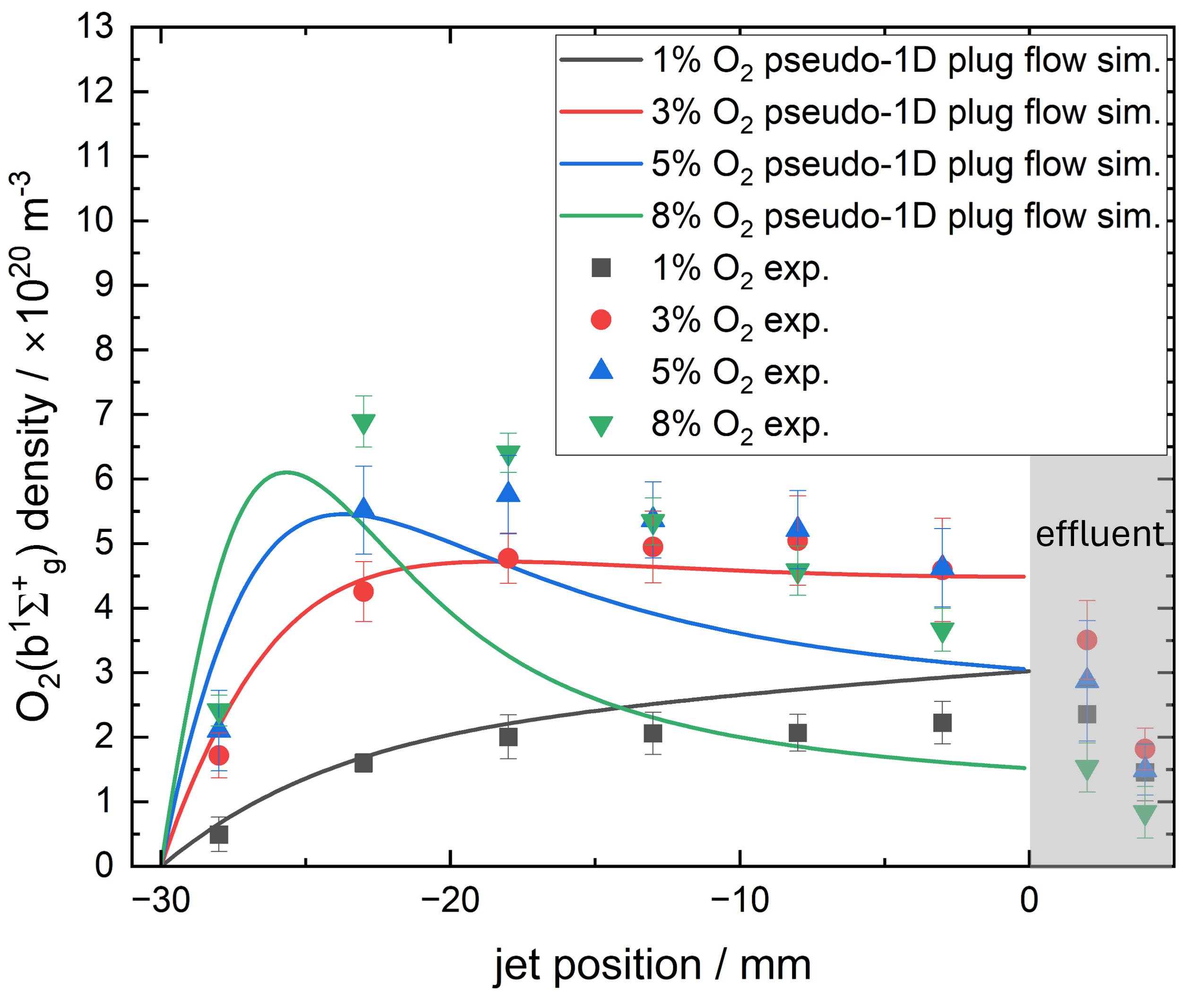}
\caption{O$_{2}(\text{b}^{1}\Sigma_{\text{g}}^{+})$ density distribution comparison along the gas channel between simulation and experiment. Grey area indicates effluent region. Example for a power of 1\,W and 1\,slm He flow and an oxygen admixture variation.\label{fig:O2b O2}}
\end{figure}

\noindent
Figure \ref{fig:O2b pos comp} compares the O$_{2}(\text{b}^{1}\Sigma_{\text{g}}^{+})$ density distribution along the COST jet discharge channel for the plug flow simulation and the experimental results. After an initial rise, the measured O$_{2}(\text{b}^{1}\Sigma_{\text{g}}^{+})$ density decreases slowly while the simulation shows a maximum at approx. -17\,mm in the discharge channel. Because the density values are heavily dependent on the position inside the jet and the oxygen admixture as shown in Figure \ref{fig:O2b O2}, we concentrated on this parameter to illustrate the differences between plug flow simulation and experimentally determined values. \\

\noindent
Figure \ref{fig:O2b O2} shows the measured O$_{2}(\text{b}^{1}\Sigma_{\text{g}}^{+})$ density distribution along the discharge channel for varying admixtures of oxygen in comparison with the 1D plug flow model. For an admixture of 0.1\,\% of oxygen the measured O$_{2}(\text{b}^{1}\Sigma_{\text{g}}^{+})$ density increases asymptotically and reaches its maximum at the end of the discharge channel, before decaying in the effluent. For an admixture of 0.3\,\% the density increases slightly and the position of the maximum is already reached at approx. -8\,mm instead at the jets exit. Afterwards it decreases slowly. The same behavior can be observed for higher admixtures. The maximum position shifts towards the gas inlet of the discharge channel and the maximum density increases slightly with higher oxygen admixture.\\

\noindent
The simulation follows the experimental values very well for 0.1\,\% and 0.3\,\% oxygen admixture. For higher admixtures a maximum in the first third of the discharge can be observed. Its position gets closer to the start of the jet with increasing admixture as observed in the experiment. The density maximum also increases with the oxygen admixture. The decay after reaching the maximum is faster than in the experimental case, which leads to lower densities at the end of the discharge channel compared to the experiment. \\

\noindent
A possible explanation for this behavior is the overestimation of the ozone density by the simulation (compare e.g. Fig. \ref{fig:O3 power}). The ozone density increases along the discharge channel to the exit of the jet leading to higher consumption of O$_{2}(\text{b}^{1}\Sigma_{\text{g}}^{+})$. As the ozone density is higher when closer to the end of the discharge the simulated O$_{2}(\text{b}^{1}\Sigma_{\text{g}}^{+})$ density reaches its maximum at the beginning of the discharge channel and decays quickly towards the end. This is consistent with the ozone density of the simulation being overestimated especially with the rising oxygen admixture as shown in Figure \ref{fig:O3 O2}. \\

\noindent
In the COST jet the O$_{2}(\text{b}^{1}\Sigma_{\text{g}}^{+})$ density predicted by the plug flow simulation has a very good agreement with the experimental values and depicts the experimental trends quite well. Higher admixtures however lead to a discrepancy in the consumption rate of O$_{2}(\text{b}^{1}\Sigma_{\text{g}}^{+})$. The overestimated consumption rate is likely caused by the overestimated ozone density as shown in the Figures \ref{fig:O3 power}, \ref{fig:O3 flow}, and \ref{fig:O3 O2}. These observations agree with the ozone measurements and confirm the suggested use cases of the two simulations. \\

\section{Conclusion}
The aim of this work was the measurement of the excited molecular oxygen species O$_{2}(\text{a}^{1}\Delta_{\text{g}})$ and O$_{2}(\text{b}^{1}\Sigma_{\text{g}}^{+})$ produced within the COST atmospheric pressure plasma jet, supplied with a helium oxygen gas mixture. The experimental results were compared to two simulations to validate the measured values and also to investigate established COST jet models for their applicability to the effluent region. \\

\noindent
Comparison of the experimental results to the pseudo-1D plug flow simulation as well as the 2D fluid simulation showed that both simulations were capable of computing the species' densities in agreement with the experiment, as both are within the same order of magnitude as the experimental densities. This makes the plug flow simulation an effective tool to compute a good estimate of the expected densities even for complicated effluent geometries. The 2D fluid simulation is able to accurately describe trends of the reactive species' densities for power, flow, and admixture variation. It is therefore well suited if the density distribution of reactive species in the effluent is critical or of special interest. Employing longer computation times to reach an equilibrium state within the flow cell might further increase the agreement with the experiment. The simulations overestimate the ozone density for oxygen admixtures higher than 0.1\% as shown in the ozone density section. This was confirmed by the O$_{2}(\text{b}^{1}\Sigma_{\text{g}}^{+})$ measurements as the consumption rate of O$_{2}(\text{b}^{1}\Sigma_{\text{g}}^{+})$ towards the jet exit was higher when compared to the experimental values. This points to a potential area of weakness in the reaction scheme used within the simulations, which might also explain discrepancies to the measured densities in the flow cell. This is consistent with the conclusions of the work of Turner \cite{turner_uncertainty_2015}, who described the close coupling between the densities of O$_{2}(\text{b}^{1}\Sigma_{\text{g}}^{+})$ and ozone under similar conditions to those studied here.  Additionally the wall reactions for the oxygen species used in the simulations are not well known under the conditions used in this work, in reality these may also deviate between the plasma and effluent regions and are even excluded for ozone. As a result, further work on improving the description of wall reactions may also improve agreement between experiment and simulation. \\

Both quantitative and qualitative differences were observed between the results of the plug flow and 2D fluid simulations. While differences in the treatment of residence time and the direct simulation of the spatial profiles of reactive species in the 2D fluid simulations are likely to be a key reason for these, the different reaction schemes used in both simulations is also likely to be a factor. In general, the reaction schemes used in each simulation are similar, but not identical. Given the sensitivity of plasma-chemical simulations to the details of certain individual reactions and the species included, even small differences can be expected to lead to noticeable changes in reactive species densities even when the physical model is unchanged, as described in \cite{turner_uncertainty_2015}, for example.\\

\noindent
To enhance the agreement of the experimental results to the simulations one could improve on the flow cell design. Reducing the appearance of vortices might reduce the simulation time for fluid models and might also lessen the discrepancy in the flow velocity considered in the plug flow simulation. Furthermore, some of the less validated rate coefficients for various reactions, particularly wall interactions, could be improved by taking a more differentiated look at the plasma and effluent region, and carefully varying those coefficients in the simulations. Nevertheless, the employed simulations showed that even outside their ``comfort zone'' they lead to a reasonably good agreement, with the experiment highlighting the quality of the plasma chemistry validation.

\section*{CRediT author statement}
\textbf{Sascha Chur:} conceptualization, methodology, formal analysis, investigation, data curation, writing -  original draft \& editing, visualization, supervision,
\textbf{Robin Minke:} methodology, validation, formal analysis, investigation,
\textbf{Youfan He:} methodology, software, validation, formal analysis, writing - original draft \& editing, 
\textbf{M\'{a}t\'{e} Vass:} methodology, software, validation, formal analysis, writing - original draft \& editing, visualization,
\textbf{Thomas Mussenbrock:} resources, supervision, project administration,
\textbf{Ralf Peter Brinkmann:} resources, supervision, project administration,
\textbf{Efe Kemaneci:} resources, supervision, project administration,
\textbf{Lars Schücke:} writing review \& editing, supervision, project administration,
\textbf{Volker Schulz-von der Gathen:} writing review \& editing, supervision,
\textbf{Andrew R. Gibson:} methodology, writing review \& editing, supervision, project administration,
\textbf{Marc Böke:} writing review \& editing, resources, supervision, project administration,
\textbf{Judith Golda:} writing review \& editing, resources, supervision, project administration.

\ack
This study was funded by the German Research Foundation (DFG) with the Collaborative Research Center CRC1316 ``Transient atmospheric plasmas: from plasmas to liquids to solids'' (projects B2, A4 and A9). The authors would like to thank Prof. Mark Kushner for providing the {\it nonPDPSIM} code.

\section{References}
\bibliographystyle{iopart-num}
\bibliography{references.bib}
\end{document}